\documentclass[preprint2]{aastex}
\usepackage{graphicx}
\usepackage{txfonts}
\usepackage{epsf,epsfig}
\usepackage{natbib}
\usepackage{subfigure}
\def\sun{\hbox{$\odot$}}

\slugcomment{For submission to ApJ}

\shorttitle{Multi-wavelength, multi-epoch angular diameters of six Mira variables.}
\shortauthors{Woodruff et al.}

\begin{document}

\title{The Keck Aperture Masking Experiment: Multi-wavelength observations of 6 Mira Variables.}

\author{H. C. Woodruff\altaffilmark{1}, 
	P. G. Tuthill\altaffilmark{1},
	 J. D. Monnier\altaffilmark{2}, 
	M. J. Ireland\altaffilmark{3},
	T. R. Bedding\altaffilmark{1}, 
	S.~Lacour\altaffilmark{1}, 
        W. C. Danchi\altaffilmark{4} , \and 
        	M. Scholz\altaffilmark{1,5}}

\altaffiltext{1}{School of Physics, University of Sydney, NSW 2006, Australia}
\altaffiltext{2}{University of Michigan at Ann Arbor, Department of Astronomy, 
                 500 Church Street, Ann Arbor, MI 48109-1090, USA}
\altaffiltext{3}{Planetary Science, Caltech, 1200 E. California Blvd, Pasadena CA 91125 USA}
\altaffiltext{4}{NASA Goddard Space Flight Center, Infrared Astrophysics, 
                 Code 685, Greenbelt, MD 20771, USA }
 \altaffiltext{5}{Institut f\"ur Theoretische Astrophysik der Universit\"at Heidelberg, Albert-Ueberle-Str. 2, 69120 Heidelberg, Germany}

\begin{abstract}
The angular diameters of six oxygen rich Mira-type long-period variables have been measured at various near-infrared (NIR)
wavelengths using the aperture masking technique in an extensive observing program from 1997 Jan to 2004 Sep.
These data sets span many pulsation cycles of the observed objects and represent the largest study of multi-wavelength, multi-epoch interferometric
angular diameter measurements on Mira stars to date.
The calibrated visibility data of $o$~Cet, R Leo, R Cas, W Hya, $\chi$ Cyg and R Hya are fitted using a uniform disk brightness distribution model
to facilitate comparison between epochs, wavelengths and with existing data and theoretical models.
The variation of angular diameter as a function of wavelength and time are studied, and
cyclic diameter variations are detected for all objects in our sample.
These variations are believed to stem from time-dependent changes of density and temperature
 (and hence varying molecular opacities) in different layers of these stars.
The similarities and differences in behaviour between these objects are analyzed and discussed in the context of existing theoretical models.
Furthermore, we present  time-dependent 3.08\,$\mu$m angular diameter measurements, probing for the first time these zones of probable dust formation,
which show unforeseen sizes and are consistently out of phase with other NIR layers shown in this study.
NIR light-curves were recovered, and show the distinctive phase lag in the maxima of $\approx$ 0.2 (compared to the visual maximum), 
similar to the lag found by, e.g. \cite{SMITH02} and \cite{NAD01}.
The S-type Mira $\chi$ Cyg exhibits significantly different behaviour compared to the M-type Miras in this study,
both in its NIR light-curves and its diameter pulsation signature.
Our data show that the NIR diameters predicted by current models are too small and need to incorporate additional and/or enhanced opacity mechanisms. 
Also, new tailored models are needed to explain the behaviour of the S-type Mira $\chi$ Cyg.
\end{abstract}


\section{Introduction}

Mira variables are pulsating M-type giants with very extended stellar atmosphere and mass-loss rates of up to $10^{-4}$M$_{\sun}/{\rm yr}$ (e.g. \citealt{JK90}).
Multi-wavelength studies allow us to probe the atmospheric structure, including H$_2$O and dust shells.
Optical and near-infrared (NIR) interferometry has been able to constrain fundamental parameters
such as intensity distributions, effective temperatures and diameters, and the dependence of these on wavelength and pulsation phase
(e.g., \citealt{HAN95}; \citealt{VANB}; \citealt{PER}; \citealt{YOU00};\citealt{HOF02}; \citealt{THOM}; \citealt{WOO}).
Interferometric studies of Mira stars conducted over multiple epochs 
and spanning a range of wavelengths have helped to address fundamental questions such
as the pulsation mode of these stars (e.g., \citealt{WOO};\citealt{FED05}), the 
molecular and dust abundances in the atmosphere (e.g. \citealt{ISTW,ISW,IS06}),
the characteristics of the circumstellar environment (e.g. \citealt{DAN94}) and photospheric/circumstellar asymmetries (e.g.  \citealt{RAG06}).\\

Interferometry, together with a host of other observational techniques, has gradually been making advances 
into our understanding of the basic physics of these stars. 
For example, there is now consensus that Miras pulsate in the fundamental mode, 
based on photometry, e.g.,  MACHO observations (see \citealt{woodM}), and spectroscopy (see \citealt{SW}) in addition to interferometry.
However, the structure and composition of the stellar atmosphere
still shows surprising results when investigated in detail.
\cite{RAG06} have, e.g.,  detected $H$ band asymmetric brightness distributions in about 29\% of their sample of nearby asymptotic giant branch stars,
substantiating another level of complexity in the structure of this class of objects.

In this paper we present the most comprehensive interferometric study of Miras to date,
encompassing 6 nearby objects, observed at up to 19 different phases in 4 filters.
With this homogeneous data-set, we are able to examine phase-dependent 
variations in Miras' atmospheric structure and are in a position to investigate long term effects that span several pulsation cycles.
By observing in the NIR (especially within the $J$, $H$ and $K$ bandpasses), we can sample molecular strata near the continuum-forming layers that 
are often close to the position of the Rosseland layer ($\tau_{\rm Ross}=1$, cf. \citealt{SCH03} and references therein).\\

Theoretical models describe the molecular layering in the atmosphere and its variation with time (e.g. \citealt{BSW}; \citealt{HSW}; \citealt{ISTW,ISW}),
and make predictions of observables such as light-curves, diameters and intensity distributions.
With our extensive database we can challenge existing models and motivate the development of a more complete physical picture of Mira variables.
Ultimately, the aim of interferometric observations like these is to calibrate the theoretical models so that fundamental parameters of Miras can be
derived from simple observables, enabling studies of stellar populations beyond the solar neighbourhood.

\section{Observations and Data Reduction}\label{observations}
\subsection{Aperture-masking Observations}

Our sample of 6 Miras contains only nearby objects, all of which have revised Hipparcos distances (see Table \ref{objects}). 
The objects were chosen for their large angular diameters and NIR brightness.
Observations were performed with the 10\,m Keck I telescope at a range of
NIR wavelengths (see Table \ref{filters}) using the Near Infrared Camera (NIRC).
The telescope pupil was converted into a sparse interferometric array by placing
aperture masks in the beam in front of the infrared secondary mirror, allowing
the recovery of the Fourier amplitudes and closure phases for baselines up
to 9.8\,m. 
For a detailed discussion of mask design, observing
methodology, scientific rationale and implementation, we refer to \cite{TUT00}. \\

For this work we used non-redundant  masks with 15 or 21 holes
configured to deliver near-optimal sampling of the Fourier plane (\citealt{GOL71}).  
Data-sets consisting of 100 140\,msec exposures were taken, alternating between the target of interest
and nearby calibrator stars. The latter were chosen to have well characterized, smaller apparent sizes (see Table \ref{calibrators}). 
The data were recorded at 19 different epochs spanning more than 7 years, delivering
good coverage through the pulsation cycles of the objects observed. 
Tables \ref{tbl-obs_ocet} to \ref{tbl-obs_rhya} list these observations.

\begin{deluxetable}{lcclc}
\tablewidth{0pt}
\tablecaption{\label{objects}
Observed objects
}
\tablehead{
\colhead{Name} & \colhead{Period} & \colhead{Spectral} &  \colhead{Distance}& \colhead{Distance}\\
			&\colhead{[days]}	&\colhead{Type Range}		& \colhead{[pc]}    & \colhead{reference}\\
}
\startdata
$o$~Cet 	&	332	& M5-9e & $107\pm6$ & 1\\
R Leo	&	312	& M6-9.5e & $82\pm5$ & 1\\
R Cas	& 	430	& M6-10e &  $100\pm5$ & 2\\
W Hya	&      385  & M7.5-9ep & $78\pm3$ & 1\\
$\chi$ Cyg&	408 & S6-S9(MSe)	&  $149\pm11$ & 1\\
R Hya	&	380 &  M6-9eS(Tc) & $118\pm7$ & 1\\
\enddata
\tablerefs{Object Period from the American Association of Variable Star Observers (AAVSO) visual light curves (A.A. Henden et al. 2006, private communication), 
M Spectral Type Range from \cite{SLOANPRICE} and $\chi$ Cyg Spectral Type from \cite{KEE80}, 
Distances from (1) \cite{KNAPP} or (2) \cite{POU02}
}
\end{deluxetable}

\begin{deluxetable}{lccc}
\tablewidth{0pt}
\tablecaption {\label{filters}  
Table of filters}
\tablehead{
 \colhead{Filter} &  \colhead{Keck/NIRC} & \colhead{Centre Wavelength } &   \colhead{Bandwidth } \\ 
 				&\colhead{Filter Name}	&	\colhead{[$\mu$m]}	&\colhead{[$\mu$m]}                  \\        
}
\startdata
z\,1.08	&HeI		&    1.083		& 0.014\\
J\,1.24 	&OII 		&    1.236		&    0.011    \\
H\,1.65	&FeII		&    1.647   	&     0.018	  \\ 
K\,2.26	&Kcont 	&    2.260 		&    0.050	 \\
		&H221	&    2.261		& 0.0239 \\
L\,3.08	&PAHcs 	&    3.082		&   0.101  	 \\
L\,3.31	&PAH	&    3.310		& 0.063	\\
\enddata

\end{deluxetable}

\begin{deluxetable}{lccc}
\tablewidth{0pt}
\tablecaption{\label{calibrators}
Calibrator stars with estimated diameters
}
\tablehead{
\colhead{Calibrator} & \colhead{Spectral} & \colhead{Adopted UD} & \colhead{Reference}\\
				&\colhead{Type} 	&\colhead{Angular Diameter [mas]} &\\
}
\startdata
$\alpha$~Cet 	&	M1.5III	 & 	$11.7\pm0.6$	&1 \\
$\alpha$ Ari 	&	K2III		 & 	$5.9\pm0.6$	& 1 \\
$\alpha$ Cas 	&	K0III	 	& 	$5.3\pm0.1$	& 2 \\
$\beta$~Cet 	&	K0III	 	& 	$6.7\pm0.5$	& 4 \\
$\alpha$ Hya 	&	K3II-III	 & 	$9.1\pm0.1$	&3 \\
$\pi$ Leo 		&	M2III		 & 	$4.6\pm0.3$	& 5 \\
$\alpha$ Lyn 	&	K7III		 & 	$7.2\pm0.6$	& 1\\
2 Cen 		&	M4.5III	 & 	$14.7$		& 6 \\
$\gamma$ Sge	&	M0III	 	& 	$6.0\pm0.6$	& 1\\
$\delta$ Sgr 	&	K3III	 	& 	$6.9\pm0.9$	& 7\\
4 Cas		&	M1III		 & 	$4.3$		& 6 \\
$\epsilon$ Cyg	&	K0III	 	& 	$4.3\pm0.4$	& 2 \\
$\delta$ Oph 	&	M0.5III	 & 	$10.1\pm0.5$	& 8\\
Vega 		&	A0V	 	& 	$3.1\pm0.1$	& 9\\
$\xi$ Cyg 		&	K4.5I	 	& 	$7.5\pm0.6$	& 10\\
$\gamma$ Hya &	G8III	 	& 	$3.0\pm0.2$	& 11\\
$\pi$ Hya 		&	K2III		 & 	$3.9\pm0.2$	& 4 \\
\enddata
\tablerefs{(1) \cite{DYCK98}; (2) \cite{MOZ91}; (3) \cite{MOZ03}; (4) from CHARM catalog \citep{CHARM};
(5) \cite{RIDG79}; (6) \cite{DUMM}; (7) \cite{MON04}; (8) \cite{PER98};
(9) \cite{HAN74}; (10) \cite{DYCK96}; (11) from CHARM2 catalog \citep{CHARM2}
}

\end{deluxetable}

\begin{deluxetable}{llllllll}
\tablewidth{0pt}
\tablecaption{\label{tbl-obs_ocet}
Observations of $o$~Cet}
\tablehead{
 \colhead{Date} & \colhead{JD}  &  \colhead{$\Phi$} &\colhead{UD$_{\rm J\,1.24}$}& \colhead{UD$_{\rm H\,1.65}$}& \colhead{UD$_{\rm K\,2.26}$}&\colhead{UD$_{\rm L\,3.08}$}&\colhead{Calibrators}\\
 			&	 \colhead{-2450000}    &				      & \colhead{	[mas] }      &  \colhead{ [mas] }           &  \colhead{[mas] }                  & \colhead{ [mas] }                  &                                      \\
 }
\startdata
1997Dec16  &   800 & 0.94	&  22.2$\pm1.0$ &27.5$^{\mbox{\tiny+2.4}}_{\mbox{\tiny-2.2}}$ &  31.0$\pm1.9$ &  58.8$\pm1.5$ &$\alpha$~Cet \\

1998Sep29 	&   1056 & 1.71	& 29.1$\pm2.2$ & 33.3$\pm1.6$&   37.0$\pm1.2$    & 57.7$\pm1.1$ & $\alpha$~Cet \\

1999Jan05 &   1184 & 2.10 	&   25.6$\pm1.3$ &27.6$^{\mbox{\tiny+2.3}}_{\mbox{\tiny-2.1}}$ & 31.8$\pm1.8$ & 61.4$\pm1.3$ &$\alpha$~Cet \\
		&		&	&		&	&		&61.5$\pm1.2$&\\
		&		&	&		&	&		&61.5$\pm1.3$&\\
		&		&	&		&	&		&61.5$\pm1.2$&\\

1999Jul29 &   1390 & 2.72	&   29.4$\pm1.9$  & 32.9$\pm1.7$ &    36.9$\pm1.2$ & 53.2$\pm1.4$ &$\alpha$~Cet\\

2000Jan25 &   1570 & 3.26 	&  &30.1$\pm2.6$ 	&  33.1$\pm1.3$ & 58.3$\pm1.4$&$\alpha$~Cet\\

2000Jun23 &   1720 & 3.71 	& 29.9$\pm1.9$ & 38.1$\pm1.8$ &        	& 56.9$\pm1.3$& $\alpha$ Ari\\
		&		&	&		&		&		&58.2$\pm1.0$&\\

2001Jul29	&   2121 & 4.92 & &30.1$\pm1.9$	&     33.8$\pm2.7$  & 58.7$\pm5.0$ &$\alpha$ Cas , $\alpha$~Cet \\

2002Jul23 &   2479 & 5.98  & 24.8$\pm2.1$ & 26.6$\pm0.5$ 	&    34.9$\pm1.5$   & 61.5$\pm2.3$&$\alpha$~Cet\\

2004Sep14 &   3262 & 8.30 & 27.1$\pm1.0$ 	& 29.3$\pm1.6$  	&    32.3$\pm1.2$    		& 59.0$\pm1.4$  &$\alpha$~Cet , $\beta$~Cet\\
		&		&	&		&			&	31.8$\pm0.9$	& 57.5$\pm1.2$ &\\
		&		&	&		&			&		& 57.4$\pm1.1$ &\\
		&		&	&		&			&		& 58.2$\pm1.9$&\\
		&		&	&		&			&		& 57.8$\pm1.1$&\\
		&		&	&		&			&		& 57.9$\pm1.4$ &\\

\enddata
\end{deluxetable}

\begin{deluxetable}{lccccccc}
\tablewidth{0pt}
\tablecaption{\label{tbl-obs_rleo}
Observations of R Leo}
\tablehead{
 \colhead{Date} & \colhead{JD}  &  \colhead{$\Phi$} &\colhead{UD$_{\rm J\,1.24}$}& \colhead{UD$_{\rm H\,1.65}$}& \colhead{UD$_{\rm K\,2.26}$}&\colhead{UD$_{\rm L\,3.08}$}&\colhead{Calibrators}\\
 			&	 \colhead{-2450000}  &				& \colhead{[mas] }        &   \colhead{[mas] }                &     \colhead{ [mas] }              &     \colhead{[mas] }               &                                      \\
 }
\startdata
1997Jan29&478&0.05&31.0$\pm1.1$ &29.6$\pm1.9$ & 30.3$\pm2.6$&52.1$\pm1.5$ & $\alpha$ Hya , $\pi$ Leo\\
	&	&		&32.6$\pm1.5$	 &  	& 32.9$^{\mbox{\tiny+2.3}}_{\mbox{\tiny-2.2}}$	&52.8$\pm1.4$  	 &\\
	&	&		&		&		&		& 55.6$\pm1.7$	&\\
	&	&		&		&		&		& 54.6$\pm1.4$	&\\
	&	&		&		&		&		&53.8$\pm1.2$	&\\
	&	&		&		&		&		&54.5$\pm1.5$	&\\
	
1997Dec16&800&1.14 &   30.0$\pm1.1$	& 		&31.4$\pm1.7$	&  47.8$\pm1.8$ 	& $\alpha$ Hya\\
	&	&		&   29.9$\pm1.2$	&		&		&		&\\

1998Apr14&918&1.54 	& 		&  32.9$\pm1.4$ 	& 32.3$\pm1.8$	& 49.2$\pm1.6$ 	&$\alpha$ Lyn\\

1998Jun04& 970 & 1.71 &  33.1$\pm2.0$&     29.7$\pm2.0$ & 32.6$\pm1.9$& 50.7$\pm1.6$ 	&$\alpha$ Lyn\\

1999Jan05&1213&2.40& 		&		&33.9$\pm0.9$	&		&$\pi$ Leo\\

1999Feb04&1213&2.49& 31.2$\pm1.8$ &   33.2$\pm2.0$ 	& 35.4$^{\mbox{\tiny+3.3}}_{\mbox{\tiny-2.9}}$ 	&  50.9$\pm1.9$	&$\alpha$ Hya\\

1999Apr25&1295&2.75& 29.6$\pm1.4$	&29.1$\pm0.8$ 	& 34.6$\pm1.3$	&	53.3$\pm1.5$	& $\alpha$ Lyn, $\pi$ Leo\\
		&	&	&		&		& 34.1$\pm1.4$	&		&\\

2000Jan25 &1570 &3.64&     	&  34.6$\pm1.6$	 &  	33.1$\pm1.7$	& 52.6$\pm0.7$	&$\alpha$ Lyn\\
		&	&	&		&		&		&52.1$\pm0.8$	&\\

2000Jun23 &1720&4.12&      	& 29.3$\pm2.1$ 	&  	31.2$\pm1.9$ & 55.4$\pm1.8$ 	&$\pi$ Leo\\

2001Jun11 &2073 &5.24& 31.5$\pm1.3$	&  33.5$^{\mbox{\tiny+4.6}}_{\mbox{\tiny-3.8}}$ 	&  30.7$^{\mbox{\tiny+4.5}}_{\mbox{\tiny-3.8}}$ 	& 56.5$\pm2.0$ 	&$\pi$ Leo ,  $\alpha$ Lyn, 2 Cen\\

2003May12&2772&7.48& 		&  36.2$\pm1.1$ 	& 37.1$^{\mbox{\tiny+5.2}}_{\mbox{\tiny-4.4}}$	&		&  $\pi$ Leo\\

2004May28&3154 &8.71&		&  30.1$^{\mbox{\tiny+2.0}}_{\mbox{\tiny-1.8}}$ 	& 32.1$\pm1.7$	& 56.4$\pm1.4$	& $\pi$ Leo\\

2005May26&3516 &9.88&  		& 28.9$\pm2.0$ 	& 31.5$\pm1.5$	&		&  $\pi$ Leo\\
		
\enddata
\end{deluxetable}

\begin{deluxetable}{lccccccc}
\tablewidth{0pt}
\tablecaption{\label{tbl-obs_rcas}
Observations of R Cas}
\tablehead{
 \colhead{Date} & \colhead{JD}  &  \colhead{$\Phi$} &\colhead{UD$_{\rm J\,1.24}$}& \colhead{UD$_{\rm H\,1.65}$}& \colhead{UD$_{\rm K\,2.26}$}&\colhead{UD$_{\rm L\,3.08}$}&\colhead{Calibrators}\\
 			&	 \colhead{-2450000}  &				& \colhead{[mas]}	         & \colhead{ [mas]}           & \colhead{[mas]}                   &  \colhead{[mas]}                   &                                      \\
 }
\startdata
1997Dec08 &802&0.67&  		&  		&30.0$\pm1.5$	&		&$\alpha$ Cas\\

1998Jun04 & 970&1.06& 25.0$\pm1.2$& 23.2$^{\mbox{\tiny+3.1}}_{\mbox{\tiny-2.7}}$ & 24.2$\pm1.6$	&44.8$\pm2.2$&$\gamma$ Sge, $\alpha$ Cas\\

1998Sep29&1056&1.27&   24.5$\pm1.5$	& 27.0$^{\mbox{\tiny+2.4}}_{\mbox{\tiny-2.2}}$ 	&		& 41.1$\pm3.0$	& $\alpha$ Cas\\

1999Jan05&1184& 1.57&   24.4$\pm1.8$	&  23.9$^{\mbox{\tiny+2.8}}_{\mbox{\tiny-2.4}}$ 	&	28.2$\pm1.2$	&	44.1$\pm1.8$	& $\alpha$ Cas\\

1999Jul29 &1390& 2.04&   24.1$\pm1.8$	&  22.4$^{\mbox{\tiny+3.4}}_{\mbox{\tiny-3.0}}$	 &	28.5$\pm1.4$	& 48.3$\pm1.5$	& $\alpha$ Cas\\

2000Jan25&1570&2.46&       		&27.7$^{\mbox{\tiny+2.2}}_{\mbox{\tiny-2.0}}$	& 29.7$\pm1.2$	&	& $\alpha$ Cas\\

2000Jun23&1720&2.81&  24.9	$\pm1.4$&     26.1$^{\mbox{\tiny+2.6}}_{\mbox{\tiny-2.3}}$ &	28.9$\pm1.9$	& 46.2$\pm1.6$ 	&$\alpha$ Cas\\

2001Jun11&2073&3.63&       	&26.5$^{\mbox{\tiny+2.4}}_{\mbox{\tiny-2.2}}$ 	&	27.4$^{\mbox{\tiny+5.4}}_{\mbox{\tiny-4.5}}$	&		& $\alpha$ Cas\\
		&	&	&		&		&	27.6$^{\mbox{\tiny+5.4}}_{\mbox{\tiny-4.5}}$	&		&\\
		&	&	&		&		&	27.7$^{\mbox{\tiny+5.2}}_{\mbox{\tiny-4.3}}$	&		&\\

2001Jul29&2121&3.74&		&  25.5$^{\mbox{\tiny+2.7}}_{\mbox{\tiny-2.4}}$ 	& 26.1$\pm2.1$ 	&		&$\alpha$ Cas\\

2002Jul23 &2479 &4.58&   		& 29.5$\pm0.9$	&	29.6$\pm1.5$	& 42.0$\pm2.0$	& $\alpha$ Cas\\
		&	&	&		&	27.4$\pm0.9$	&		&		&\\

2003May12&2772&5.26&       	&23.3$\pm2.3$ 	& 	22.1$^{\mbox{\tiny+14.3}}_{\mbox{\tiny-7.7}}$	& 		&$\delta$ Sgr\\
		&	&	&		&	24.4$\pm1.4$	&	29.0$^{\mbox{\tiny+8.0}}_{\mbox{\tiny-6.0}}$	&		&\\

2004May28&3154&6.14&  	&  23.2$^{\mbox{\tiny+2.9}}_{\mbox{\tiny-2.5}}$ 	&  24.6$^{\mbox{\tiny+2.2}}_{\mbox{\tiny-2.0}}$	& 48.9$\pm1.5$	&4 Cas\\

2004Sep2&3262&6.39&27.6$\pm0.9$	&26.8$^{\mbox{\tiny+1.8}}_{\mbox{\tiny-1.6}}$&25.2$\pm1.5$&42.2$\pm1.3$& $\beta$~Cet , $\alpha$ Cas ,$\epsilon$ Cyg \\
		&	&	& 26.6$\pm1.1$	&	 26.5$\pm1.8$	&	24.5$\pm1.3$	&	40.5$\pm2.1$	&\\
		&	&	&		&	27.9$\pm2.0$	&		&		&\\

\enddata
\end{deluxetable}

\begin{deluxetable}{lccccccc}
\tablewidth{0pt}
\tablecaption{\label{tbl-obs_whya}
Observations of W Hya}
\tablehead{
 \colhead{Date} & \colhead{JD}  &  \colhead{$\Phi$} &\colhead{UD$_{\rm J\,1.24}$}& \colhead{UD$_{\rm H\,1.65}$}& \colhead{UD$_{\rm K\,2.26}$}&\colhead{UD$_{\rm L\,3.08}$}&\colhead{Calibrators}\\
 			&	 \colhead{-2450000}    &				      &	 \colhead{[mas]}      &   \colhead{[mas]}            &  \colhead{[mas]}                   &  \colhead{[mas]}                  &                                      \\
 }
\startdata
1997Jan29&478&0.53&41.9$^{\mbox{\tiny+3.2}}_{\mbox{\tiny-3.2}}$&38.9$\pm1.8$& 42.4$\pm1.8$& 62.3$\pm1.4$& $\delta$ Oph\\
&&&42.0$^{\mbox{\tiny+4.1}}_{\mbox{\tiny-4.2}}$&38.3$\pm1.0$&42.5$\pm2.2$& 61.7$\pm1.1$&\\
	&	&		&		&		& 41.9$\pm1.8$& 60.3$\pm1.0$&\\
	&	&		&		&		& 39.4$\pm2.0$& 60.6$\pm1.5$&\\
	&	&		&		&		& 39.7$\pm2.7$&&\\
	&	&		&		&		& 41.4$\pm2.2$&&\\

1998Apr14&918&1.76&    32.8$\pm2.4$  & 34.2$\pm1.7$& 37.4$\pm1.8$& 62.3$\pm1.2$& $\alpha$ Lyn, 2 Cen\\
	&&	&	&	34.6$\pm2.0$&39.8$\pm1.4$&&\\
	&&	&	&				&36.7$\pm1.2$&&\\

1998Jun04&970&1.90& 32.0$\pm2.3$&    33.6$\pm1.8$ & 36.4$\pm$1.1& 62.7$\pm1.0$& 2 Cen\\

1999Jan05&1184& 1.50&		& 				&43.0$\pm1.4$ &&2 Cen\\
&&&		& 				&41.8 $\pm1.2$      &&\\
&&&		& 				&      42.4$\pm1.7$       &&\\
&&&		& 				&42.6 $\pm1.8$      &&\\

1999Feb05&1213&2.58& 39.3$\pm3.0$&   41.6$\pm1.2$& 42.0$^{\mbox{\tiny+2.7}}_{\mbox{\tiny-2.5}}$& 63.2$\pm1.2$& 2 Cen\\
	      &       	&	&		&	&43.8$^{\mbox{\tiny+2.4}}_{\mbox{\tiny-2.2}}$&&\\

1999Apr25&1295&2.79&  35.2$\pm3.5$&     38.1$\pm2.4$& 41.0$\pm1.0$&&2 Cen\\
	&	&		&		& 36.7$\pm2.2$&36.6$\pm1.6$&&\\
	&	&		&		& 36.3$\pm1.7$&&&\\
	&	&		&		& 36.8$\pm1.8$&&&\\
	
1999Jul29&1390&3.04&     		&  34.8$\pm1.6$&36.7$\pm0.9$ && 2 Cen\\

2000Jan25&1570&3.53& 47.1$\pm2.7$&  43.2$\pm1.5$ & & 64.5$\pm1.1$& 2 Cen\\
		&		&	&			&			&45.9$\pm0.8$&			&\\

2000Jun23 &1720&3.94&		& 35.0$\pm1.8$ 	&		& 68.5$\pm1.6$	& 2 Cen\\

2001Jun11&2073&4.88&  34.5$\pm3.7$&    36.0$\pm1.8$& 41.7$^{\mbox{\tiny+3.1}}_{\mbox{\tiny-2.9}}$& 65.3$\pm1.4$&  $\alpha$ Lyn, 2 Cen\\
	&	&		&		&		& 39.4$^{\mbox{\tiny+3.1}}_{\mbox{\tiny-2.9}}$&&\\

2003May12&2772&6.66& 		&	40.6$\pm1.3$& 44.2$^{\mbox{\tiny+3.8}}_{\mbox{\tiny-3.3}}$ &&  2 Cen\\

2004May28&3154&7.56&       	& 39.1$\pm1.5$ & 41.5$\pm1.4$& 65.7$\pm1.2$& 2 Cen\\
&	&		&		&		&41.4$\pm1.2$&&\\

2005May26&3516&8.55&       	&38.4$\pm1.9$ &43.8$\pm1.4$ && 2 Cen\\
		
\enddata
\end{deluxetable}

\begin{deluxetable}{lccccccc}
\tablewidth{0pt}
\tablecaption{\label{tbl-obs_xcyg}
Observations of $\chi$ Cyg}
\tablehead{
 \colhead{Date} & \colhead{JD}  &  \colhead{$\Phi$} &\colhead{UD$_{\rm J\,1.24}$}& \colhead{UD$_{\rm H\,1.65}$}& \colhead{UD$_{\rm K\,2.26}$}&\colhead{UD$_{\rm L\,3.08}$}&\colhead{Calibrators}\\
 			&	 \colhead{-2450000}    &				      & \colhead{[mas]}       &   \colhead{[mas]}            &  \colhead{[mas]}                   &  \colhead{[mas]}                   &                                      \\
 }
\startdata

1998Jun04&970 0&0.59&  21.3$\pm2.8$ &    19.5$^{\mbox{\tiny+3.9}}_{\mbox{\tiny-3.1}}$ & 23.0$\pm2.1$&  45.6$\pm3.0$& $\gamma$ Sge\\

1998Sep29&1056&0.87&		&      21.2$^{\mbox{\tiny+3.4}}_{\mbox{\tiny-2.9}}$ & 26.1$\pm2.7$& 46.0$\pm4.0$& $\gamma$ Sge\\

1999Jul29&1390&1.62& 21.6$\pm1.6$&     22.4$^{\mbox{\tiny+3.2}}_{\mbox{\tiny-2.7}}$ & 28.2$\pm1.5$& 42.4$\pm3.0$& $\gamma$ Sge, Vega\\
	&		&	&		&		&		& 43.7$\pm2.8$&\\
	
2000Jun23&1720&2.48&		&     23.1$^{\mbox{\tiny+3.0}}_{\mbox{\tiny-2.6}}$ & 25.1$\pm2.2$& 39.5$\pm2.4$&$\gamma$ Sge, $\xi$ Cyg\\
	&		&	&		&		& 24.9$^{\mbox{\tiny+2.3}}_{\mbox{\tiny-2.1}}$& 41.0$\pm2.7$&\\
	
2001Jun11&2073&3.36& 25.4$\pm1.9$&      26.4$^{\mbox{\tiny+2.8}}_{\mbox{\tiny-2.5}}$& 30.4$^{\mbox{\tiny+4.7}}_{\mbox{\tiny-4.1}}$& 45.5$\pm2.5$& $\gamma$ Sge\\

2002Jul23&2479&4.36&  28.5$\pm2.7$&     27.5$\pm1.0$ & 29.6$\pm1.6$& 44.3$\pm2.5$& $\gamma$ Sge\\

2003May12&2772&5.07&		&     28.1$\pm2.0$& 42.1$^{\mbox{\tiny+4.4}}_{\mbox{\tiny-3.9}}$& 46.5$\pm2.6$& $\epsilon$ Cyg\\

2004May28&3154&6.01&		&     24.7$^{\mbox{\tiny+3.0}}_{\mbox{\tiny-2.7}}$& 33.8$\pm3.5$ & 53.1$\pm4.0$& Vega\\

2004Sep23&3262&6.31& 24.8$\pm1.2$&      26.7$\pm2.0$ & 33.0$\pm2.3$&  50.5$\pm3.9$& $\gamma$ Sge\\

\enddata
\end{deluxetable}

\begin{deluxetable}{lccccccc}
\tablewidth{0pt}
\tablecaption{\label{tbl-obs_rhya}
Observations of R Hya}
\tablehead{
 \colhead{Date} & \colhead{JD}  &  \colhead{$\Phi$} &\colhead{UD$_{\rm J\,1.24}$}& \colhead{UD$_{\rm H\,1.65}$}& \colhead{UD$_{\rm K\,2.26}$}&\colhead{UD$_{\rm L\,3.08}$}&\colhead{Calibrators}\\
 			&	 \colhead{-2450000}    &				      &\colhead{[mas]}       &   \colhead{[mas]}            &  \colhead{[mas]}                   &  \colhead{[mas]}                   &                                      \\
 }
\startdata
1997Jan29&478&0.70&   26.2$\pm1.9$&&  26.0$^{\mbox{\tiny+2.8}}_{\mbox{\tiny-2.5}}$ & 40.0$\pm2.1$& $\delta$ Oph\\

1999Jan05&1184&2.56&		&		&  24.3$\pm2.3$ & & $\gamma$ Hya\\
	&		&	&		&		& 27.7$\pm0.9$&&\\
	&		&	&		&		&	27.4$\pm1.0$&&\\
	&		&	&		&		&	24.0$\pm1.9$&&\\

1999Feb05&1213&2.64&		&      30.0$^{\mbox{\tiny+2.2}}_{\mbox{\tiny-2.0}}$&&& $\pi$ Hya\\

1999Apr25&1295&2.86&       	& 24.1$\pm2.9$&28.0$\pm2.2$&& 2 Cen\\

2000Jan25&1570&3.58&      	&		& 31.0$\pm1.3$ && 2 Cen\\

2003May28&2772&6.74&       	&		&27.2$^{\mbox{\tiny+9.0}}_{\mbox{\tiny-6.3}}$ & & 2 Cen\\

\enddata
\end{deluxetable}

\subsection{Extraction of Visibilities}\label{dr}

The procedures for extracting the visibility amplitudes, as well as
engineering  and performance details, are documented in 
\cite{MON99t} and \cite{TUT00}, while recent scientific applications
of the data pipeline can be found in \cite{MON02} and \cite{TUT02}.
In principle, the pupil geometry of the telescope mimics the operation of a separate-element
 interferometer array, and the data collection and analysis are similar to standard  methods for interferometry
 experiments such as speckle imaging.
 The short-exposure images are dark-subtracted, flat-fielded and cleaned of pattern noise.
Power spectra are then computed frame by frame as the squared modulus of the Fourier transform.
 Stellar fringes appear as discrete peaks in such power spectra, with the origin occupied by a peak 
 whose height is proportional to the squared flux in the frame.
 Squared visibilities are found by dividing the power at the spatial frequency of the fringes by
  that at the origin and then normalizing with the corresponding signal from the calibrator spectrum.
 The uncertainty associated with the squared visibilities is derived from the scatter in each ensemble of 
 100 exposures.\\
 
 \subsection{Seeing Correction}\label{seco}
 
\begin{figure*}[htbp]
\begin{center}
\epsscale{2.2}
\plotone{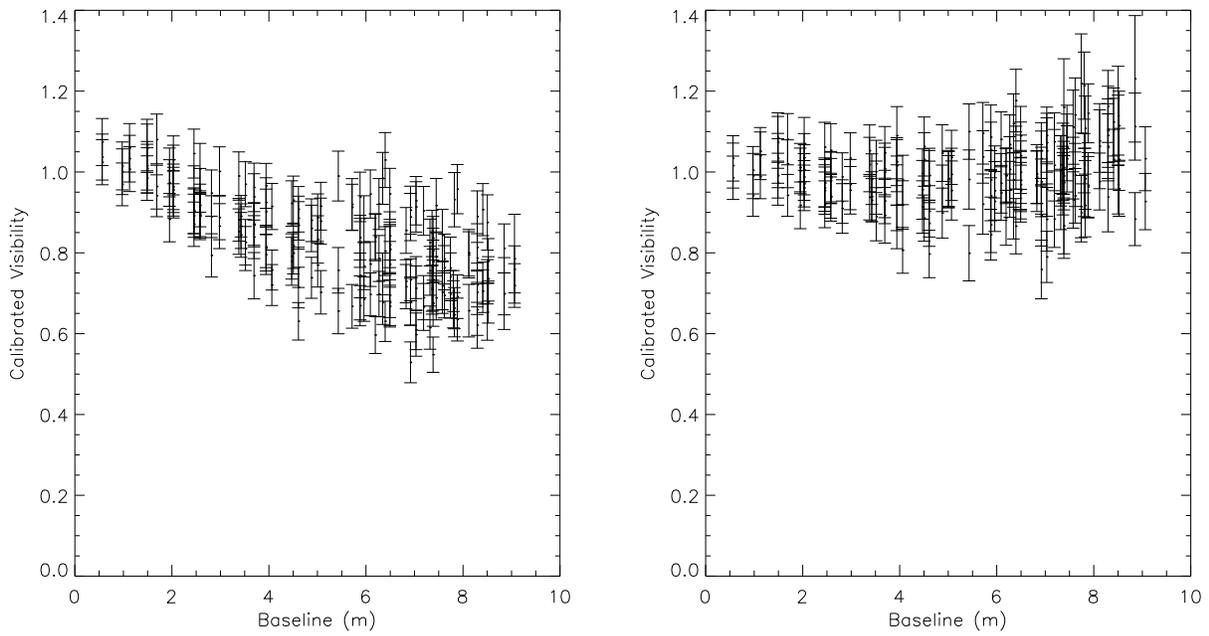}
\end{center}
\caption{Left: Calibrated visibilities (without azimuthal averaging) of an unresolved target star ($\alpha$ Ari). 
Due to the effects of variable seeing and/or windshake, the object appears to have an UD angular diameter of 23\,mas.
Right: The same data with the seeing correction applied, showing that these data are now consistent with an unresolved source. }
\label{windshake}
\end{figure*}

For all baselines and for all target and calibrator stars considered in
this paper, the dominant noise sources were seeing and windshake (wind-induced telescope wobble). 
When the atmospheric conditions vary between observing the source and its calibrator, the overall ratio
changes between the fringe power and the total flux on the detector. 
As clarified by \cite{MON04}, for aperture-masking data this change is
nearly constant as a function of baseline for baselines longer
than the coherence length ( $\approx$ 0.5 m at K\,2.26 band).
This means the visibility function will approach a non-unity value at short baselines.
We took the visibility at the origin to be a free parameter in our studies, an assumption that does not
affect our fitting procedures as long as there is no significant flux coming from over-resolved structures ($>$ 0".5).
Seeing and windshake also lowered the mean visibility V and increased its variance on each baseline.
 However, the quantity $\frac{{\rm Var}(V)}{<V^2>}$ was independent of the stellar brightness and
the degree to which it was resolved by the baseline in question, only depending on seeing and windshake.
We partially corrected for the effects of seeing and windshake using the method described by \cite{IRE2006},
which entails empirically fitting a function of the form $V = V_0 \exp(-k \frac{{\rm Var}(V)}{<V^2>})$ to
individual star observations and correcting visibilities for each target
and calibrator star prior to dividing the target with calibrator visibilities. 
We used a conservative value for $k$ of 0.8 (which may 
have under-corrected for the effects of seeing and windshake) and smoothed
the function $\frac{{\rm Var}(V)}{<V^2>}$ in the $u-v$ plane in order to minimise
errors in applying this correction (for further details see \citealt{IRE2006}).
We evaluated the effectiveness of this correction by comparing the calibrated visibilities
of unresolved stars with and without seeing correction, which turned out to improve the calibration process.
The apparent angular diameters were closer to the true values after the seeing/windshake correction
on $>70\%$ of the stars examined (see Figure \ref{windshake}).
Only on 3 object observations with very bad windshake did the algorithm
cause a marginal worsening of the miscalibration (5\,mas).

\begin{figure*}[htbp]
\begin{center}
\epsscale{2.3}

\plottwo{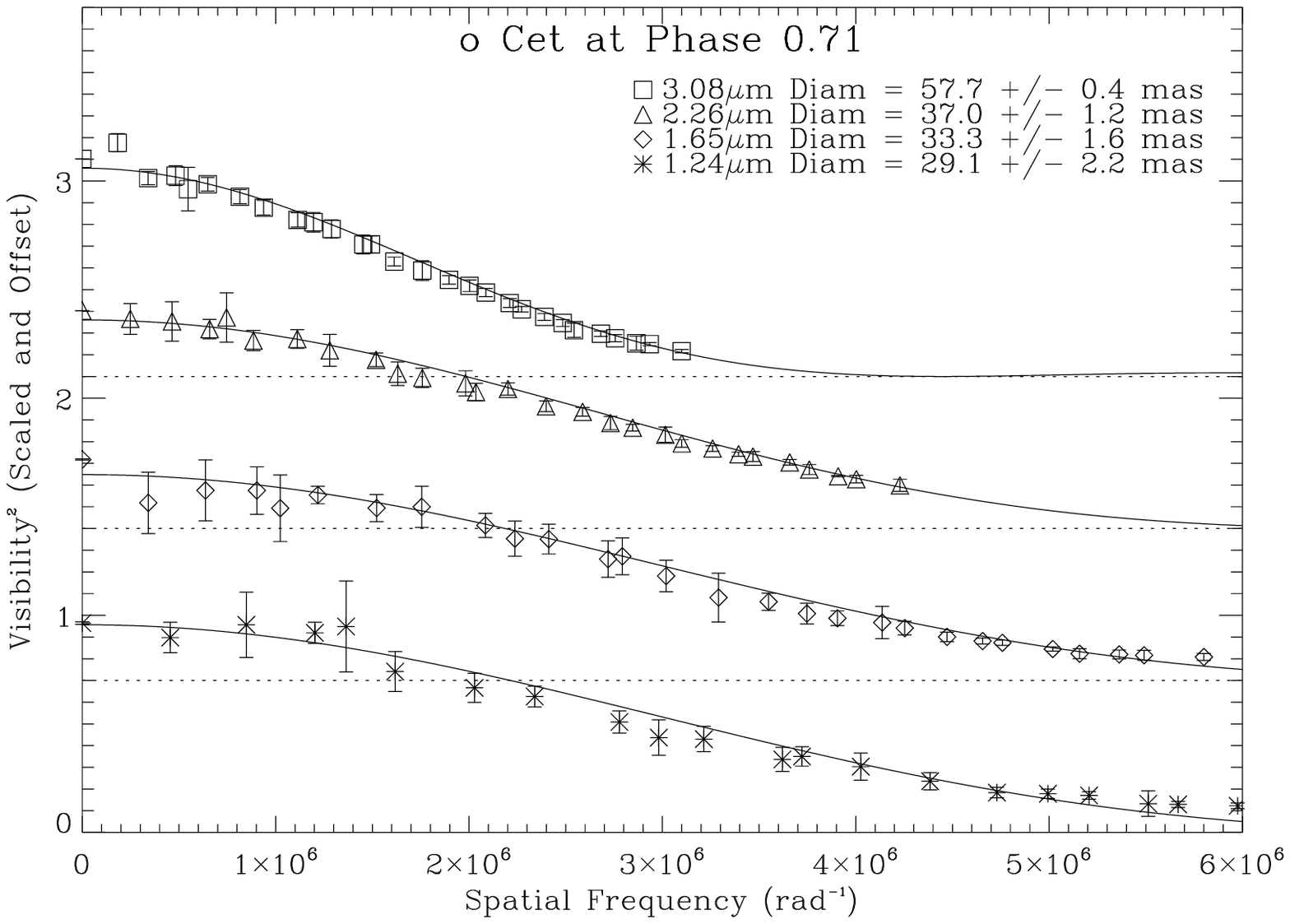} {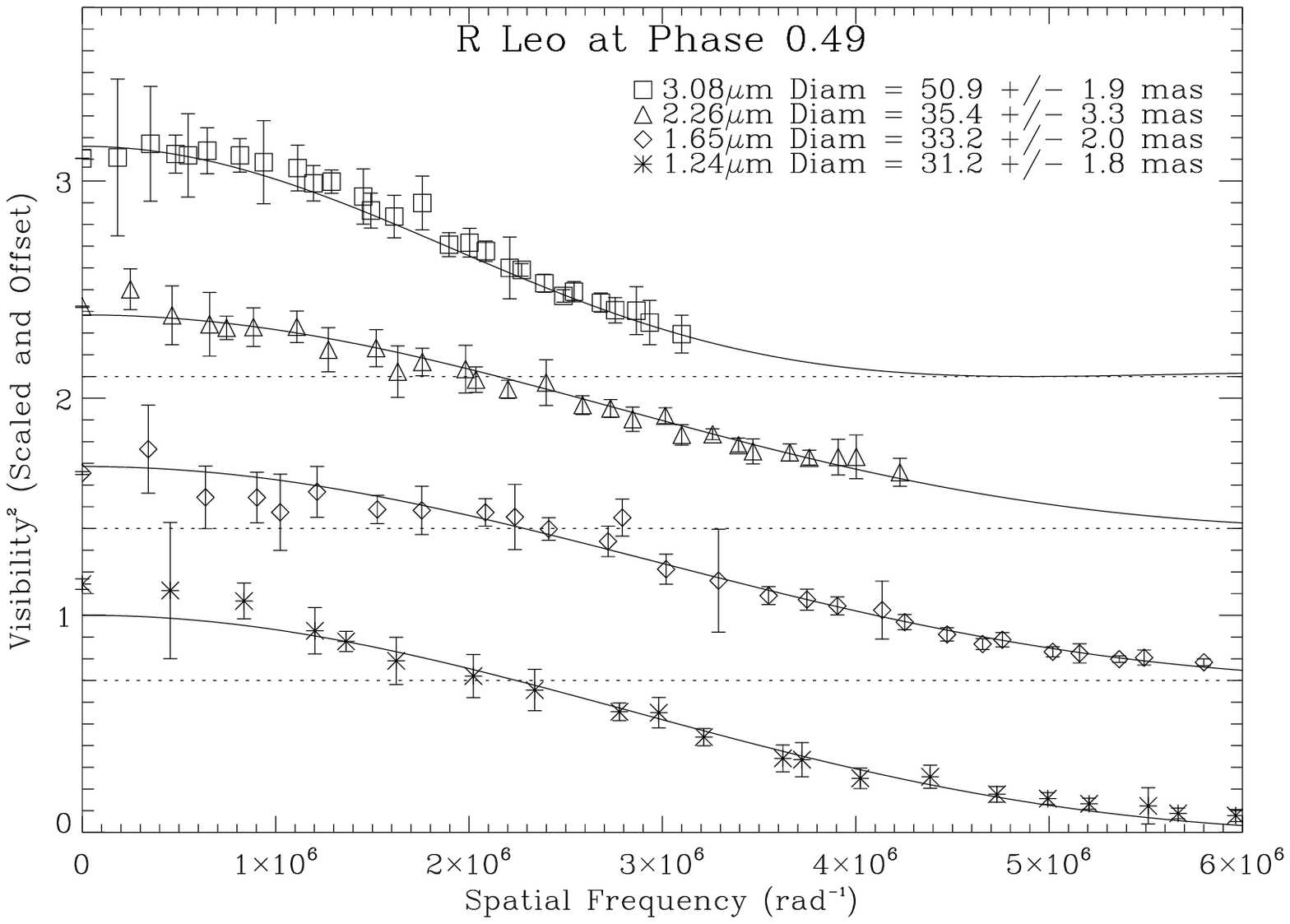}

\plottwo{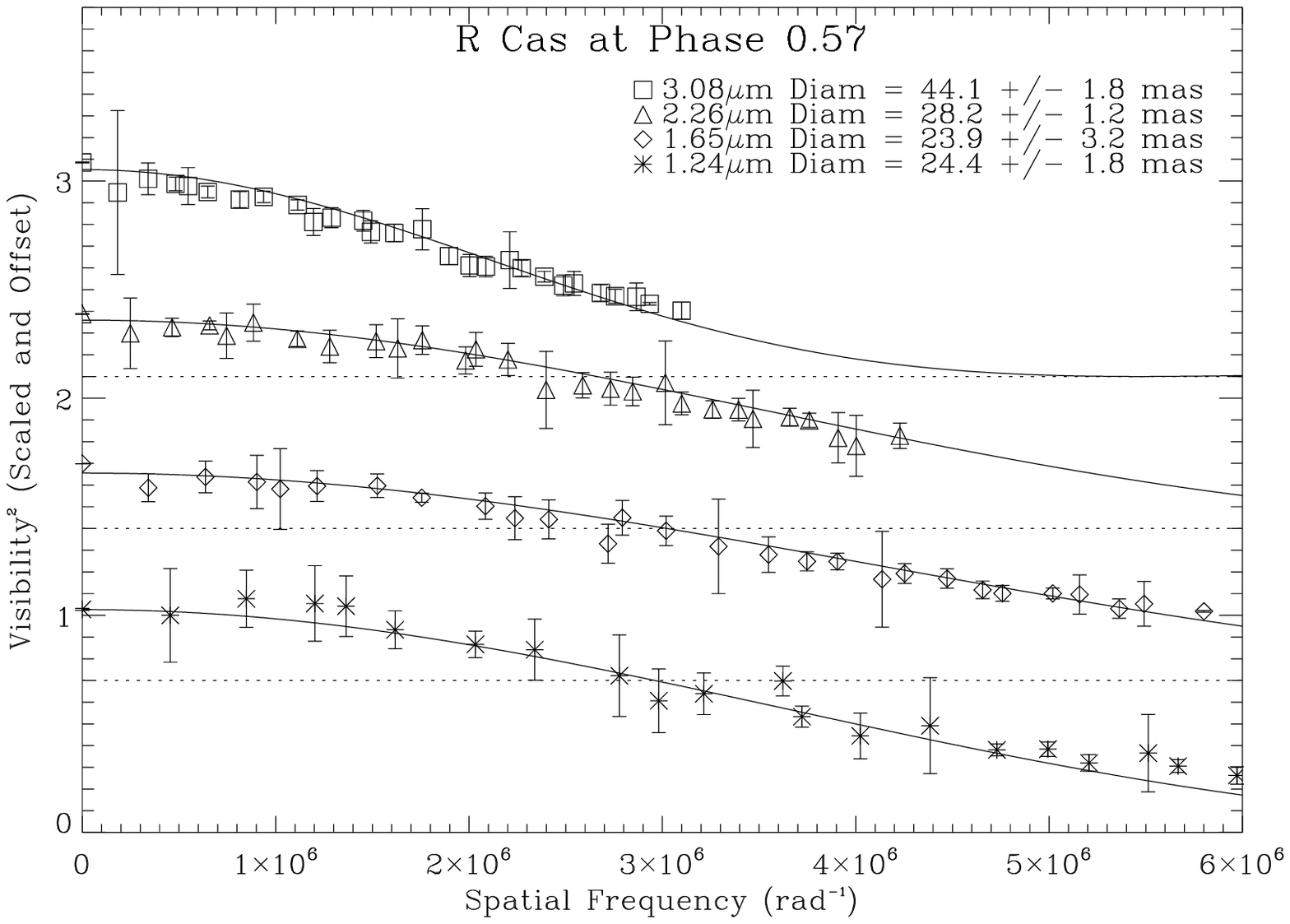} {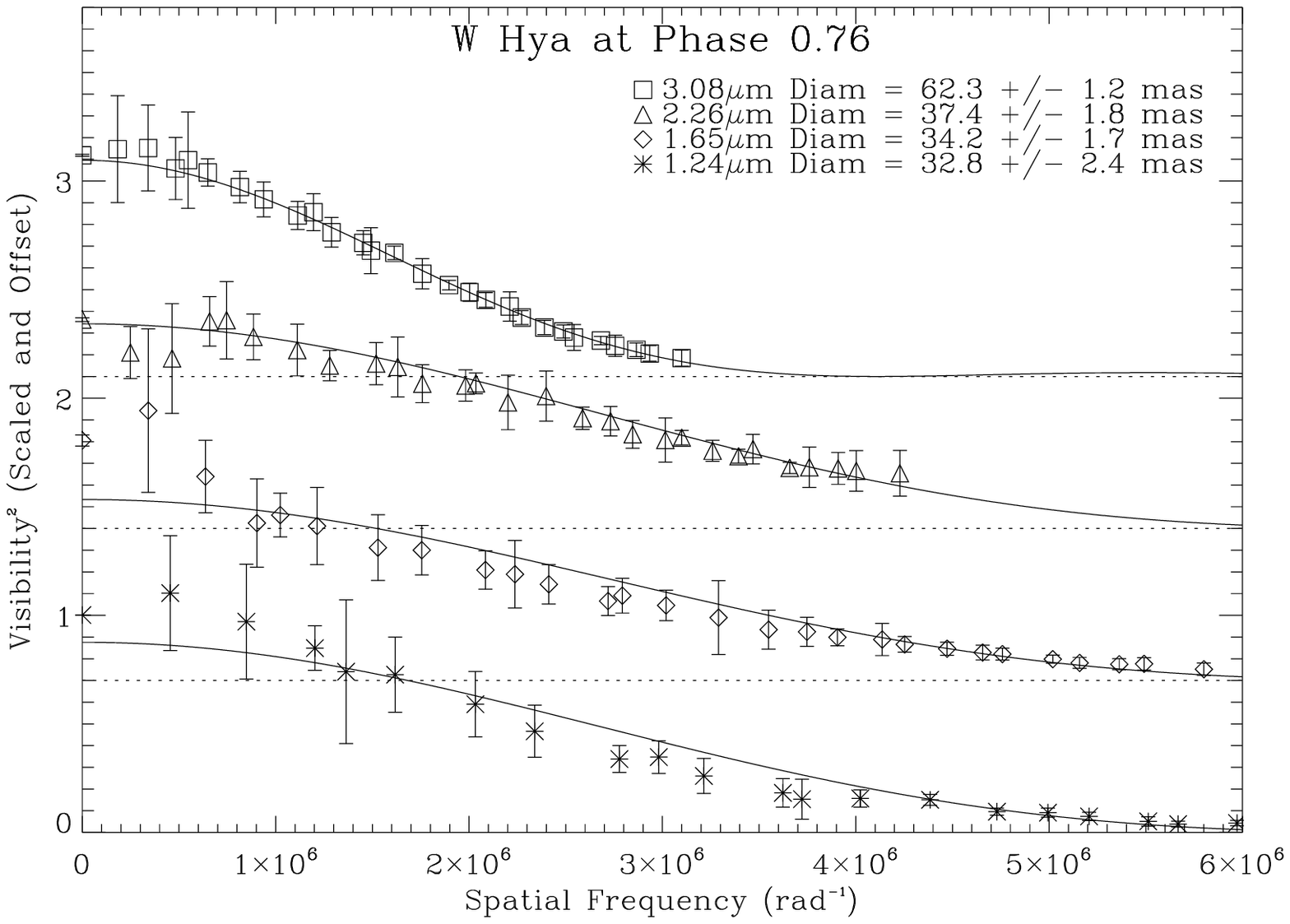}

\plottwo{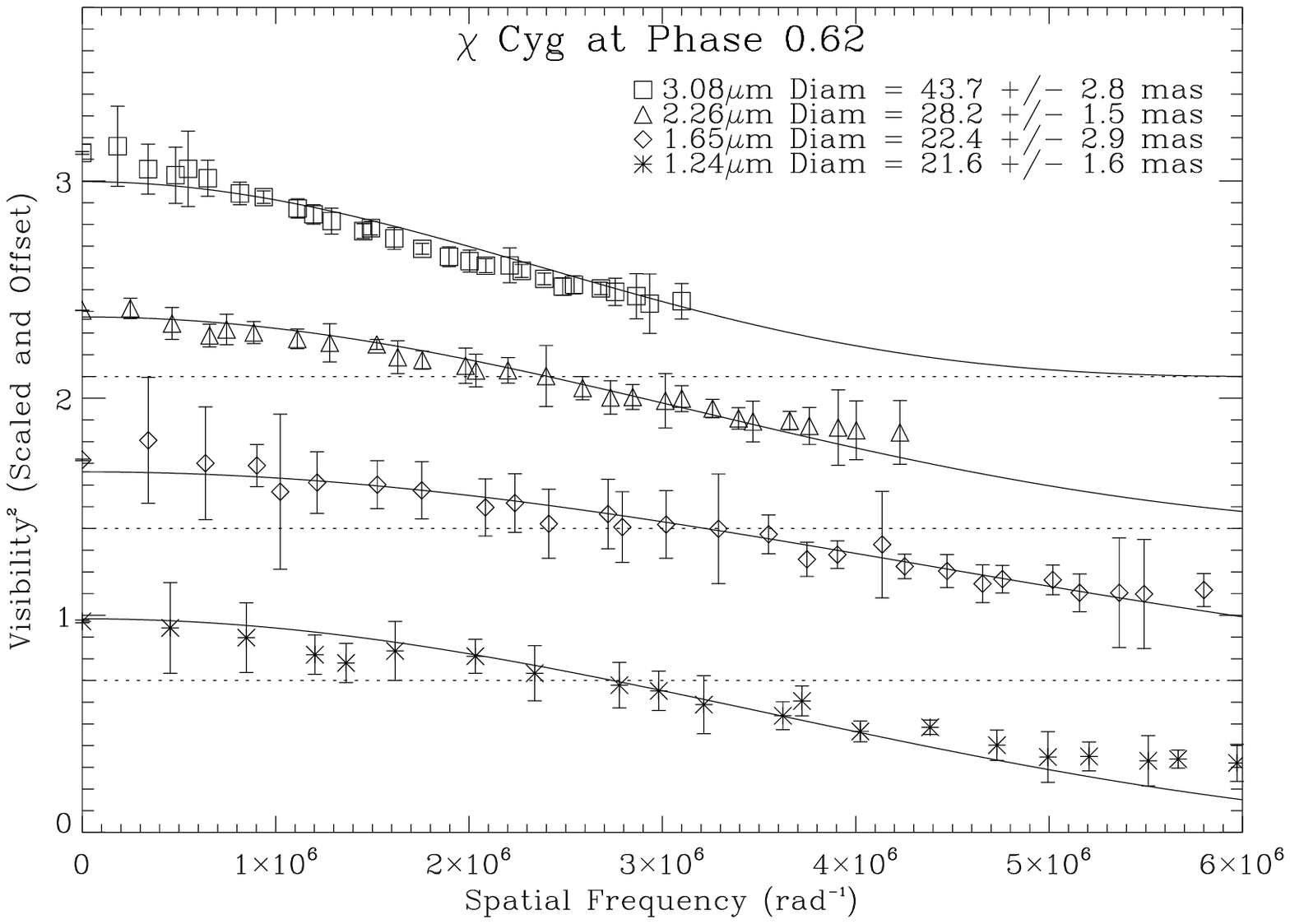} {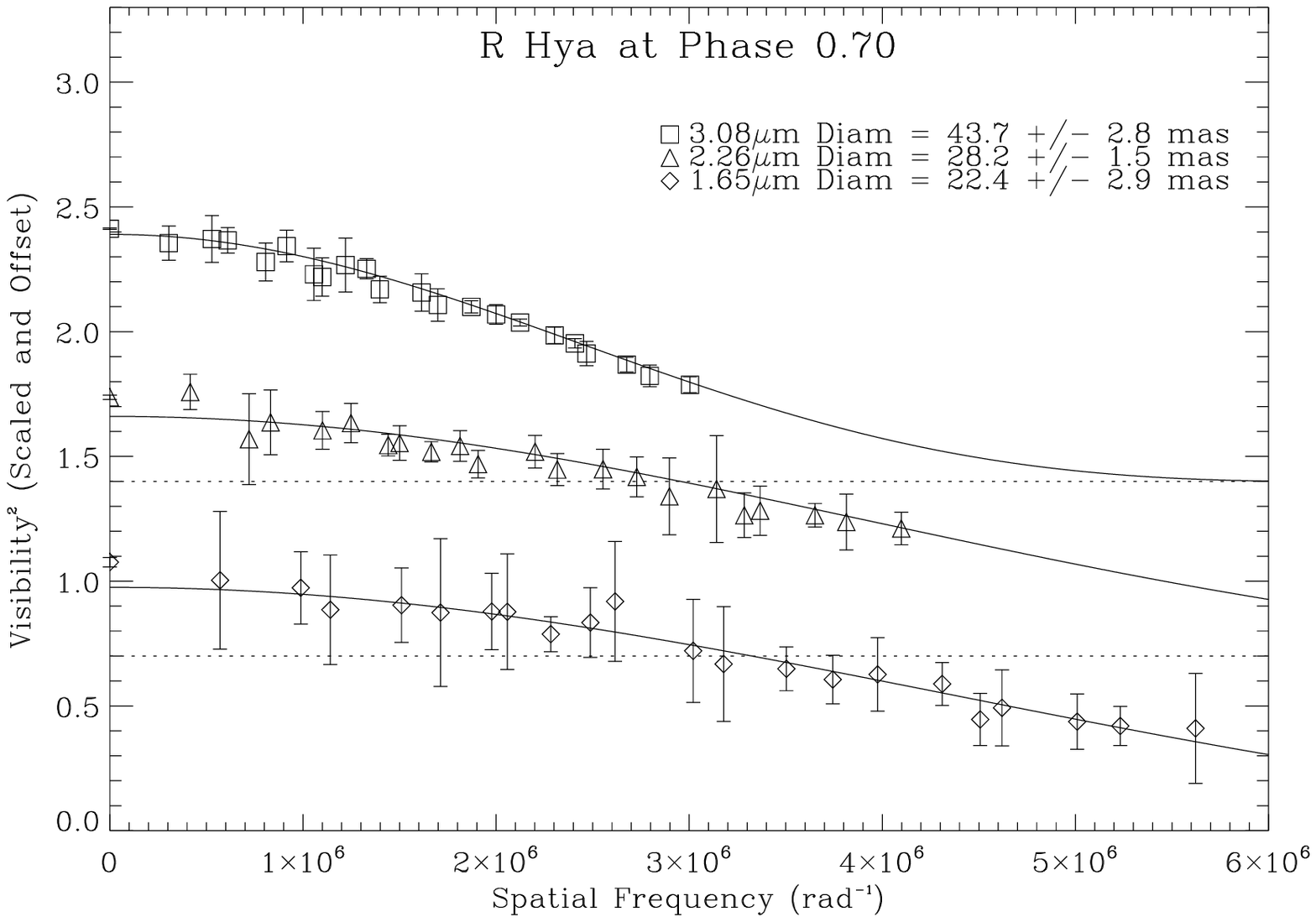}

\end{center}
\caption{Azimuthally averaged squared visibility functions for $o$~Cet, R Leo, R Cas, W Hya, $\chi$ Cyg and R Hya at various visual phases
Results for the different filters are offset for clarity.
The very large apparent diameters in the L\,3.08 filter can be attributed mainly to the H$_2$O opacity, as first suggested and observed by \cite{TUT00b}. The diameters correspond to best-fit UD models.}
\label{typical}
\end{figure*}

Figure \ref{typical} shows typical data obtained from the masking experiment. 
The targets were seldom fully resolved,
making image reconstruction only possible for those objects with the largest angular diameters (e.g. W Hya). 
These results will be presented in a subsequent paper.
The two dimensional visibilities showed no significant deviation from circular symmetry (except for W Hya), possibly due to
the lack of high spatial frequencies sampled by our $<$10\,m baselines.
Since W Hya's  deviation from circular symmetry is smaller than the UD angular diameter error,
we made the assumption that all the stars are spherical and the data were azimuthally averaged.

To keep the representation of the data homogeneous, our quantitative analysis is based on fitting simple uniform disk (UD)
models to the azimuthally averaged Fourier data.
We concentrate on the four filters which provide most of the temporal coverage in our observations, namely
L\,3.08, K\,2.26, H\,1.65 and J\,1.24, considering the other filters only in Section \ref{atmos}.
Here, e.g., L\,3.08 indicates that the filter covers a sub-interval of the L bandpass of the conventional UBV filter system
and is centered at 3.08\,$\mu$m (Table \ref{filters}). Note that the K\,2.26 bandpass encompasses two very similar filters,
both centered at 2.26\,$\mu$m with slightly different bandwidths.

Calibrator stars nearby in the sky and measured interleaved
with the target observation are used to estimate the system transfer function,
a standard practise in interferometry described in detail by \cite{MG_MIRA_05}, 

The visibility amplitude was calibrated by dividing the target visibilities
by the calibrator star visibilities, after first correcting for the estimated sizes of the calibrators (Table \ref{calibrators}).

\subsection{Uniform Disk diameters}

Although the true stellar intensity profile is not a UD, fitting the observed visibilities
 with this simple profile still provides a useful estimate of the apparent size of the target.
Our longest baseline was just under 10\,m, and so we are resolving low-resolution structure in the target star's intensity profile.
This makes it difficult to differentiate between a UD, a fully-darkened disk, a Gaussian or a more complex intensity distribution. 
We chose UD diameters to allow comparison of findings with existing literature and to avoid the difficulties encountered when deriving more sophisticated diameters (cf. \citealt{HSW};\citealt{SCH03}).
If the ``true'' intensity distribution is known, or a predicted model distribution is to be compared with the data, different radius definitions can be easily converted from and to a UD radius.

In various stars, the J\,1.24-band curves show some deviation from the UD model at around $3\times10^6$\,rad$^{-1}$, which could
indicate the presence of dust causing scattering in the outer layers, leading to a smaller "true" photospheric diameter (see Figure \ref{typical})
than derived with a simple UD model fit. 
In addition, due to their larger angular diameters (especially in the L\,3.08 filter), W Hya and possibly $o$~Cet were also resolved enough to 
detect further deviations of the intensity distribution from the simple UD shape to be compared with model predictions.
 They are also possible targets for imaging after recovering the closure-phase information
in addition to the Fourier amplitudes.
These comparisons with more complex models and interpretations will be addressed in a future paper.
 
\subsection{Estimation of Seeing Miscalibration}\label{error}

Given that the seeing correction described in Section \ref{seco} leaves some residual error and 
assuming that the dominant error term is the uncertainty introduced by variations in the seeing between measurements of targets and calibrators, 
we undertook a separate study which assessed the robustness and repeatability of the calibration process to seeing-induced miscalibration.
If the seeing remains constant, then calibrating (i.e., dividing) the visibility function from successive datasets taken on the same object
should yield the visibility function of a point source (i.e V$^2 =1$). 
In the case of changing atmospheric conditions, considerable differences in the transfer function can appear between consecutive datasets.
The error induced by such fluctuations can be measured from observations of a single object 
(either target or calibrator) taken through the same filter at different times during one night.\\
Once again, we chose to fit a UD model to the data, a natural candidate to fit the error introduced by the change in seeing throughout one observing night.
This allows us to estimate atmospheric uncertainties by applying the same fitting procedure as was used for the rest of our data,
reducing the risk of biases introduced by the fitting method.

Asymmetric errors were calculated by calibrating the target with a UD of the derived diameter.
These errors pertain to the respective filters, and if no object was observed twice in one night in the same band,
the maximum global error is assumed.

The data in Figure \ref{typical}  demonstrate another well known problem: that of calibration near the so-called seeing spike. 
Note how the azimuthally averaged visibility points at short baselines  can deviate considerably from the otherwise uniform shape of the visibility curve.
As the calibration of these large fluctuations at low spatial frequencies is highly challenging,  
we fitted our models only to the spatial frequencies in Fourier space corresponding to baselines of 2m or more.

\subsection{NIR photometry}\label{lc}
   
In the course of the data reduction, we extracted the total received flux for both science objects and calibrators.
With this we were able to retrieve contemporaneous photometry data in the the same bands as 
our angular diameter measurements.
These measurements complement the photometry by \cite{WHI}, which does not cover our
full range of observations, and provide a powerful tool to monitor NIR light-curves.
The small differences between the \cite{WHI} photometry and our measurements can be attributed to our use of filters of differing bandpass and center wavelength(see Table \ref{filters}).

\begin{figure*}

\epsscale{2.5}
\plottwo{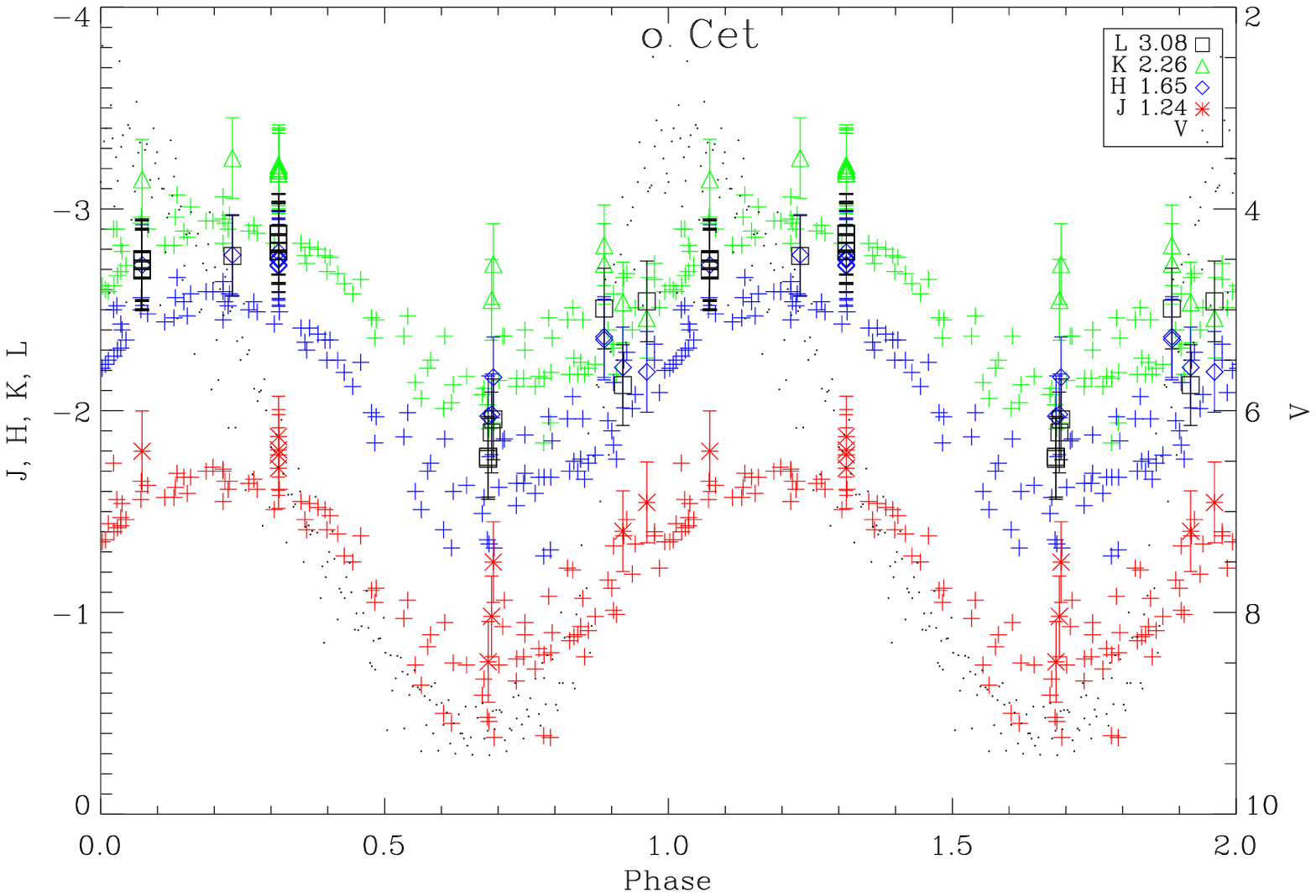}{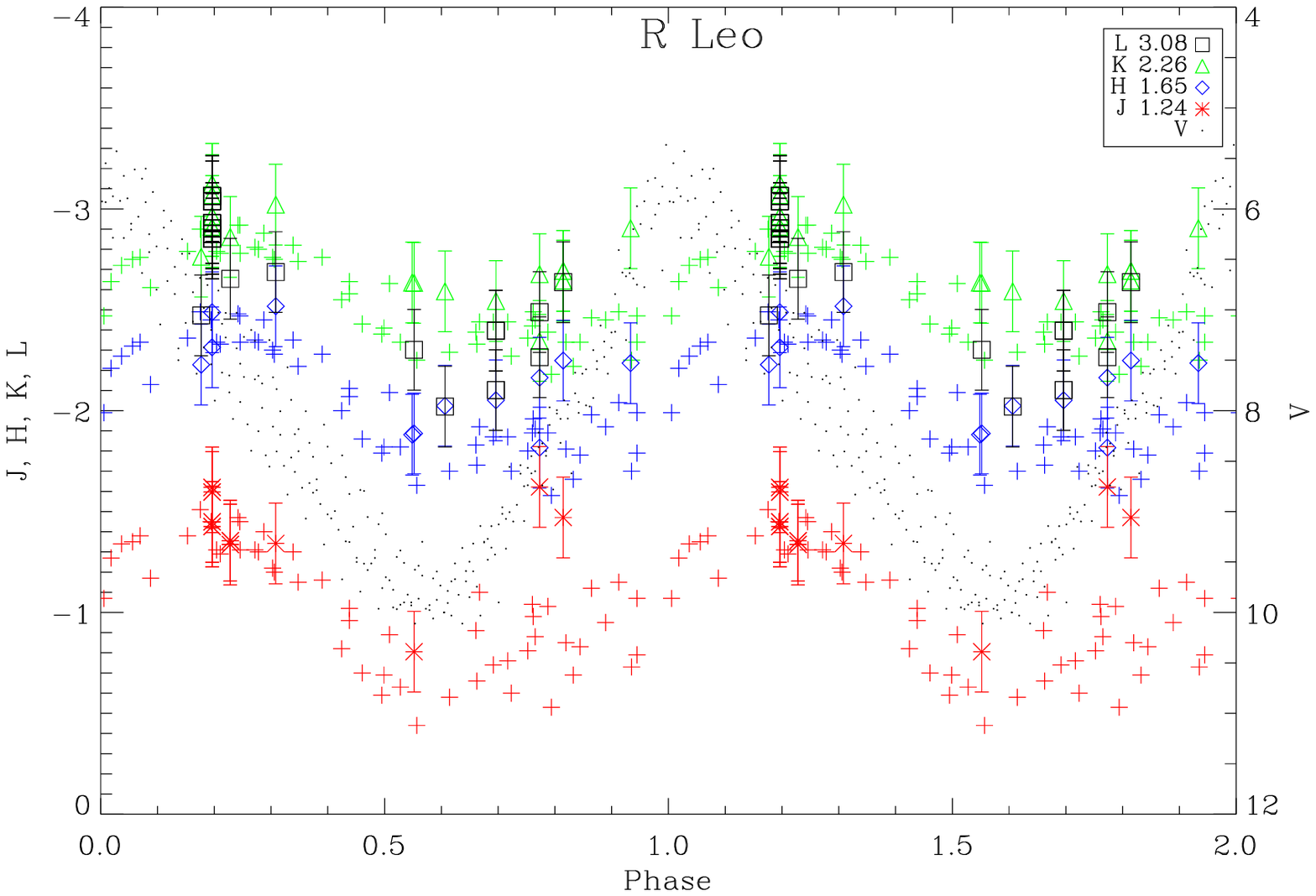}
\plottwo{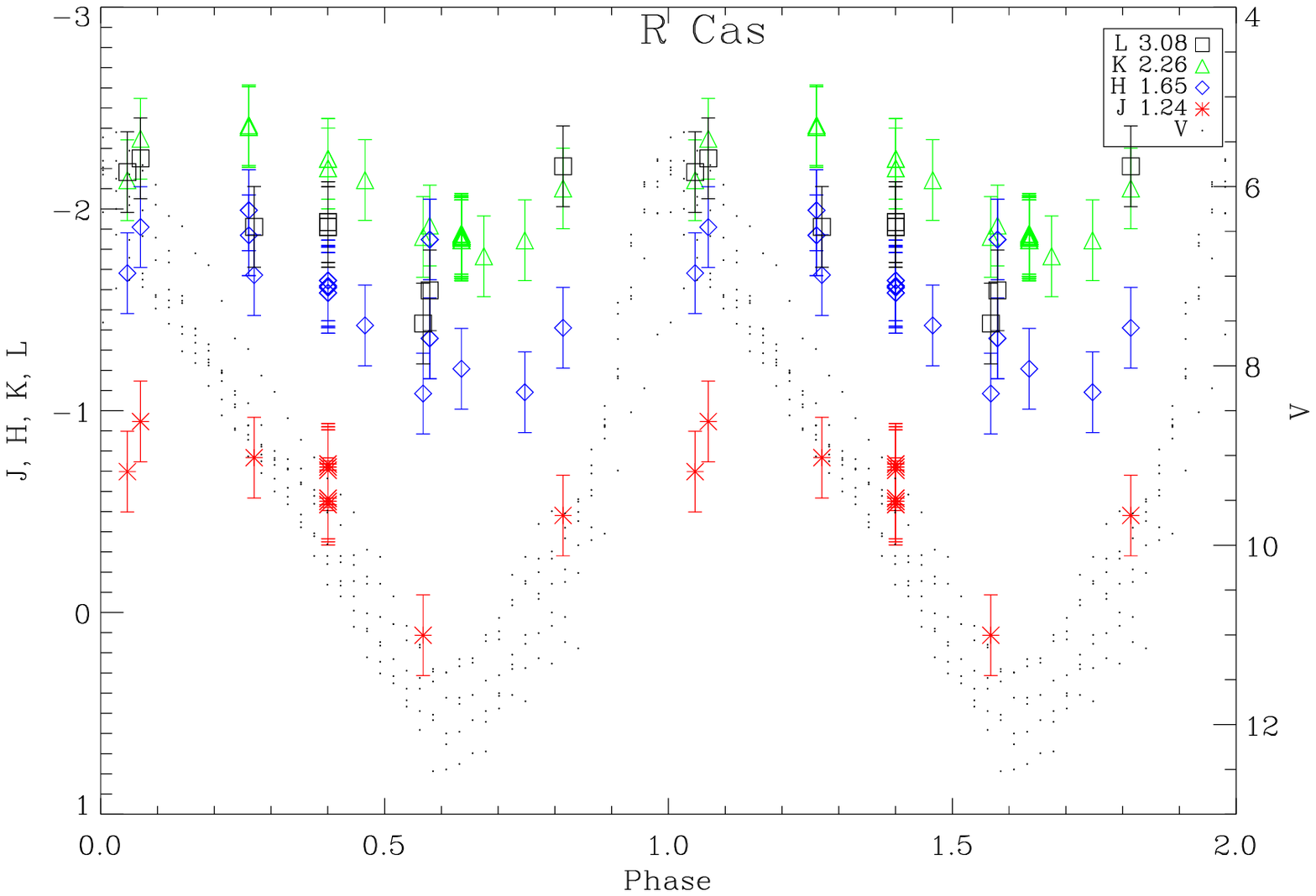}{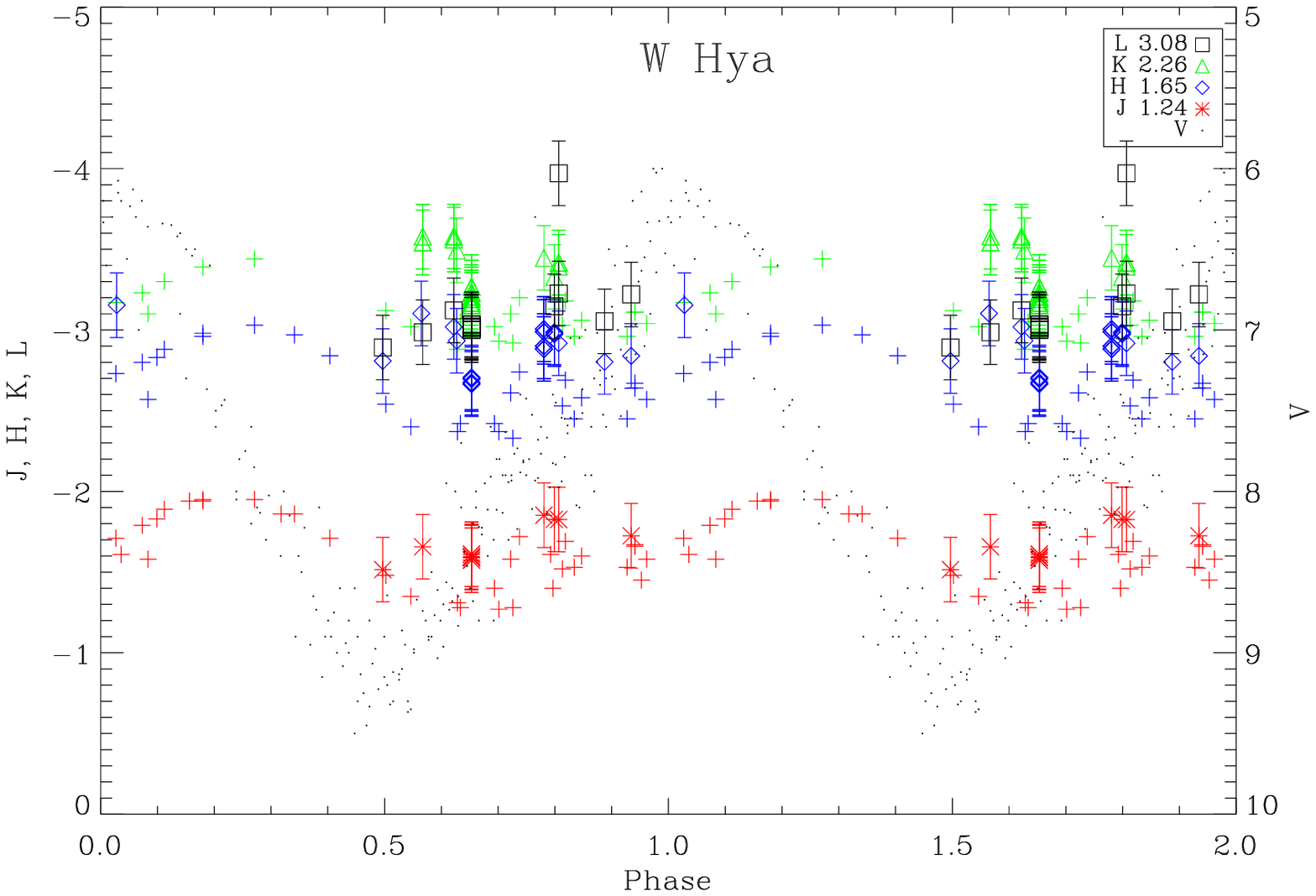}
\plottwo{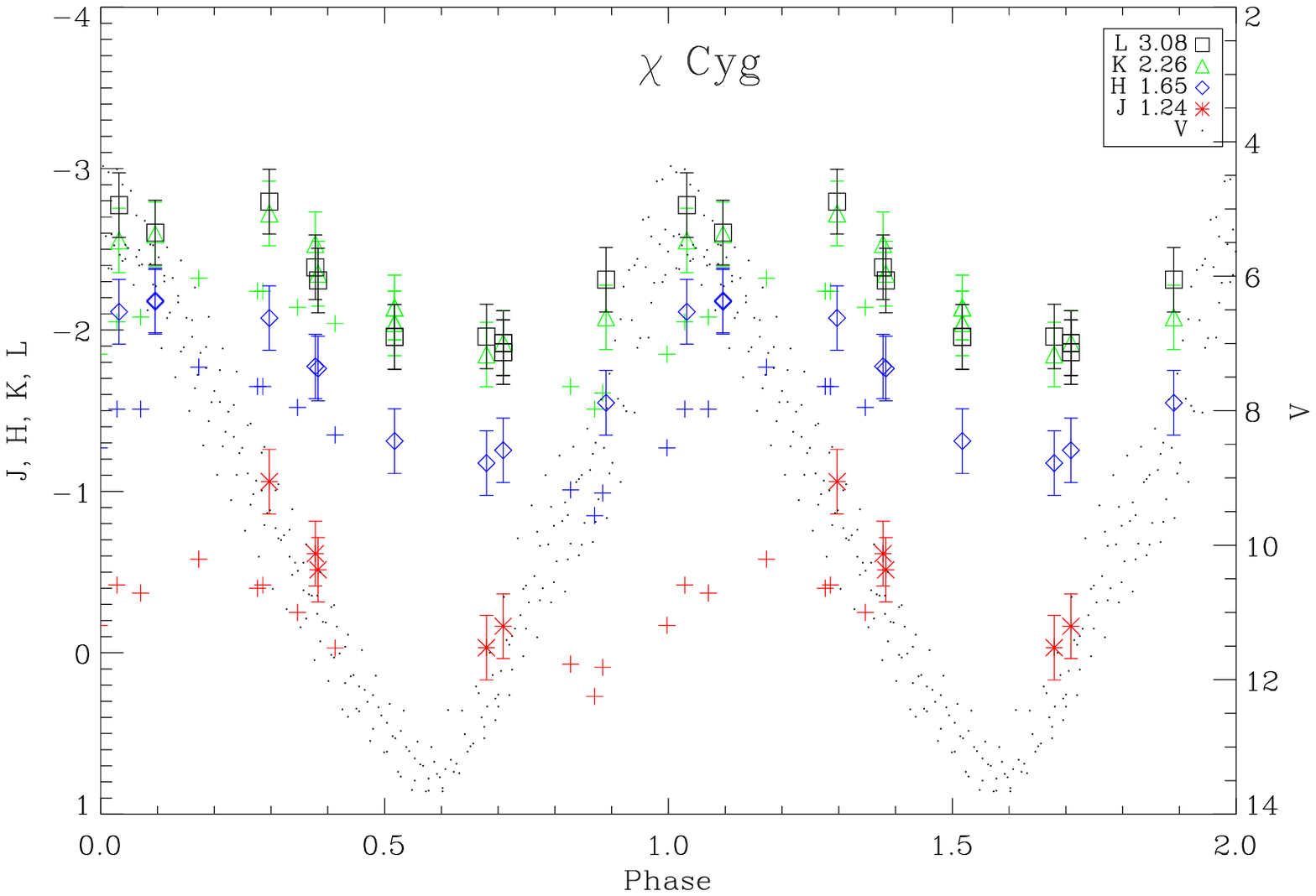}{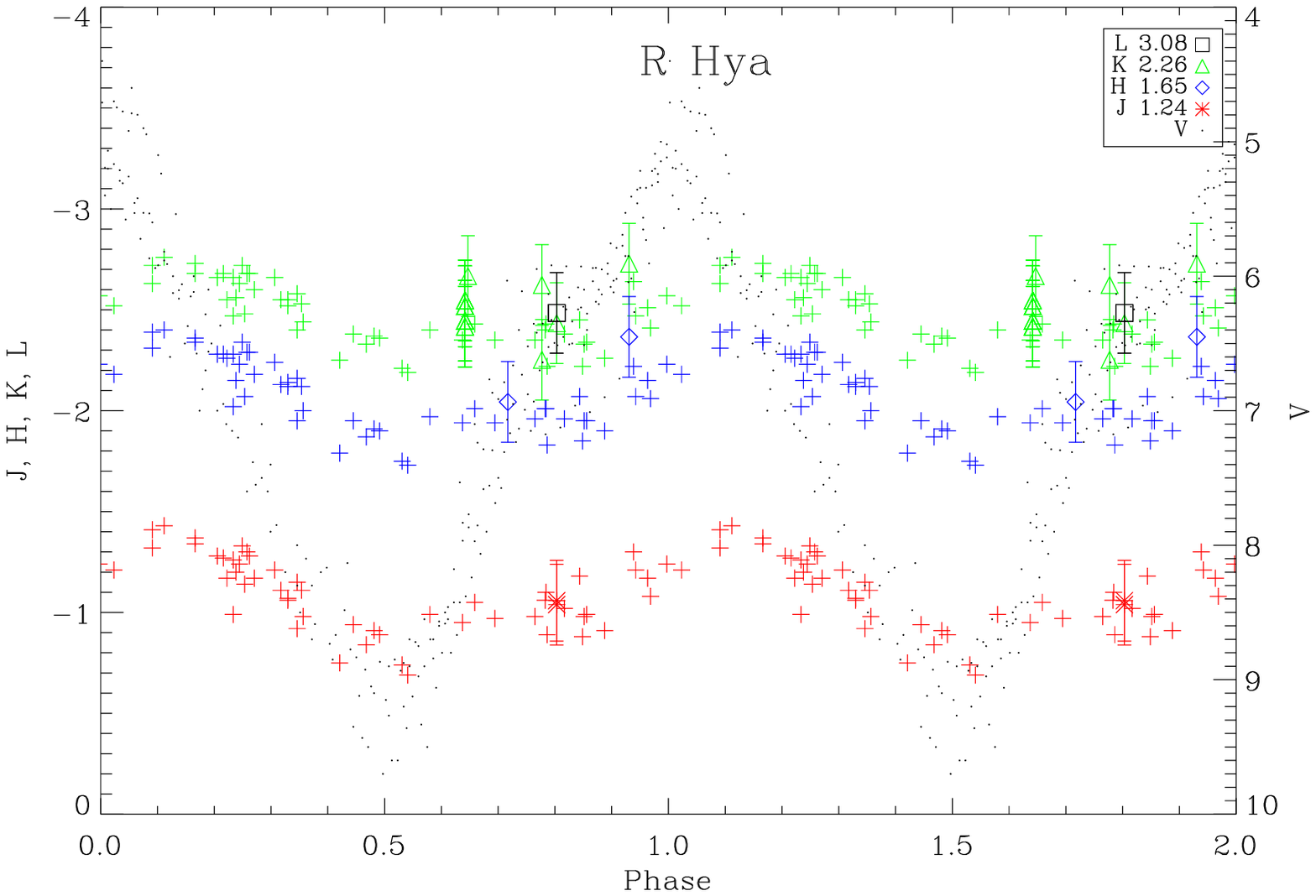}

\caption{All Figures: Our NIR photometry folded with the visual pulsation period, shown with \cite{WHI} $JHK$ photometry and visual light-curves. 
The J\,1.24,H\,1.65,K\,2.26 and L\,3.08 fluxes extracted from Nirc masking are represented by red asterisks, blue diamonds, green triangles and black squares respectively.
\citealt{WHI} J (red crosses), H (green crosses) and K photometry data (red crosses) are shown where available. Visual data kindly supplied by the AAVSO. }
\label{jhk_whitelock_nirc}
\end{figure*}

\section{Results}

\subsection{Light-curves}\label{lightc}

Figure \ref{jhk_whitelock_nirc} shows our NIR photometry plotted as a function of pulsation phase, together with the $JHK$ photometry
of \cite{WHI} and AAVSO visual photometry (A.A. Henden et al. 2006, private communication).
As an approximation based on the appearance of the light curves, we fitted our NIR photometry data with simple sinusoidal functions (not shown in the figures), 
with the exception of R Hya, where the light curve was too incomplete.
Although these fits are only approximations, they provide a useful tool to examine the dates of maxima and minima as a function of wavelength.
The NIR magnitude vs. phase curves are certainly not strictly sinusoidal, 
but depending on the star and the bandpass they can be approximated by a sine function.
Our photometry shows a small lag of up to 0.07 cycles in the maxima of light curves from the J\,1.24 to the H\,1.65 and from the H\,1.65 to the K\,2.26 filters,
as well as less pronounced or non-existing lags near visual minimum,
confirming the findings of \cite{SMITH02}, as can be seen in Figure \ref{jhk_whitelock_nirc}.
The reported  phase shift of $\approx0.15-0.22$ with which the NIR maxima lag behind 
visual maxima typical for M-type Miras  (\citealt{NAD01}, \citealt{SMITH02}) is also clearly seen in our data.
Also in accordance with the findings of \cite{SMITH02},  we find that the maxima in longer wavelengths (i.e. in the $L$ band) 
occur before the $K$ maxima but after the visual maxima for all of our objects except for W Hya, where the $L$ light curve is too peculiar
to be fitted with a sine function.\\

Both Visible and NIR lightcurves given in Figure \ref{jhk_whitelock_nirc}
span the observation period of each object and  clearly illustrate the substantial differences 
between the cycle amplitudes and pulsation periods.

\subsection{Multi-wavelength, multi-epoch UD angular diameters}

\begin{figure*}[htbp]
\epsscale{1.7}
\plotone{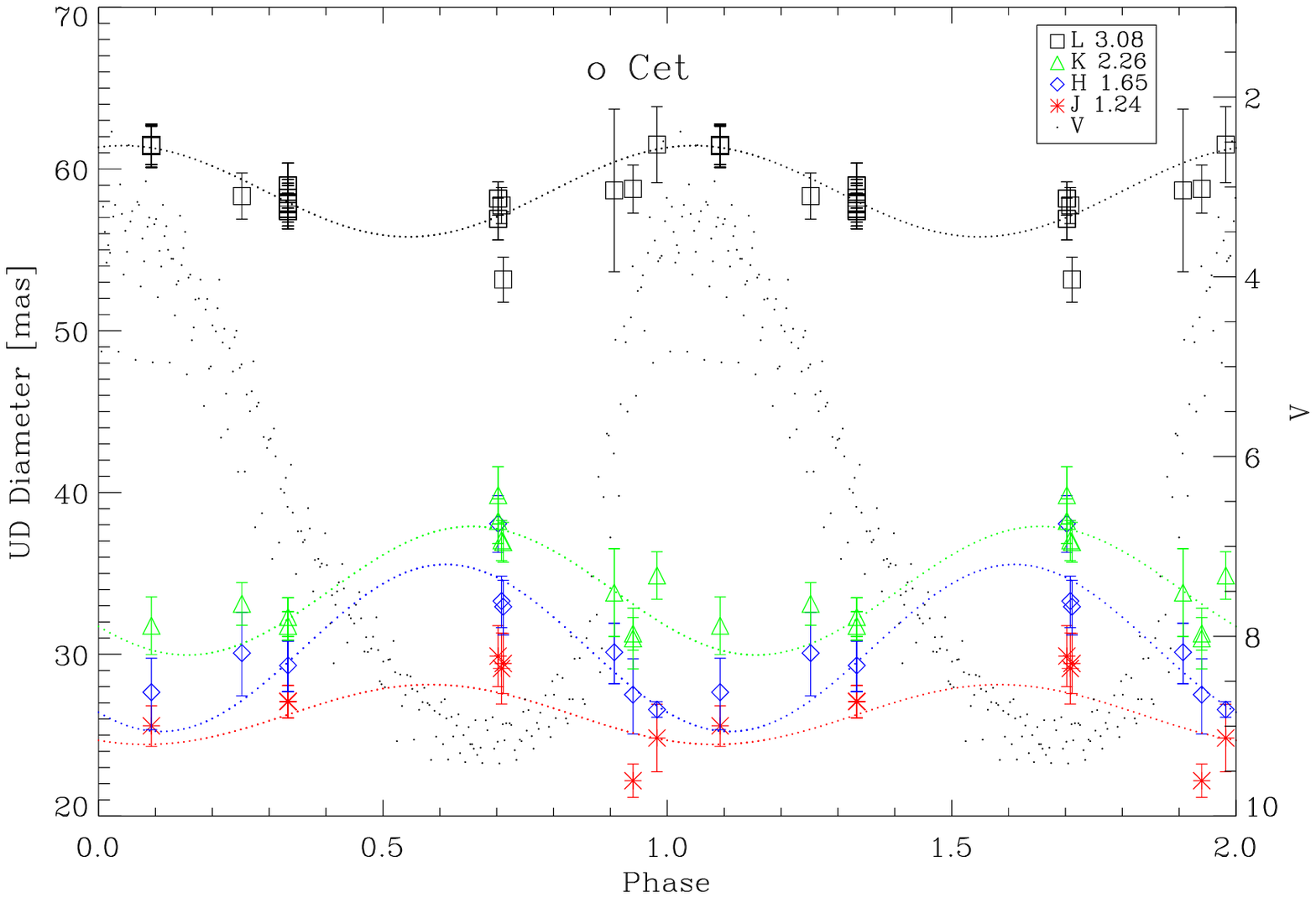}
\plotone{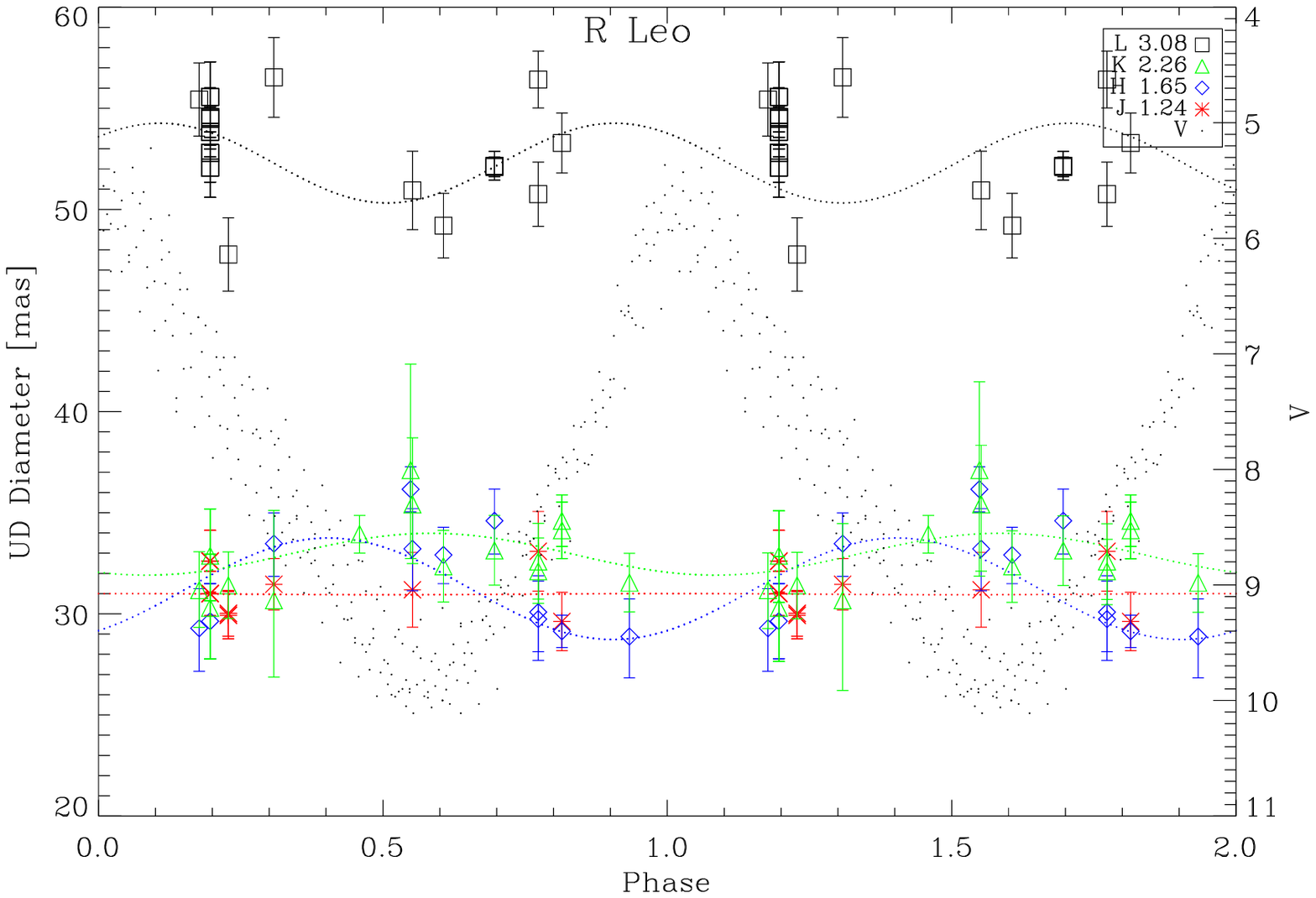}

\caption{
Uniform disk angular diameters for $o$~Cet (upper panel) and R Leo (lower panel), folded with pulsation phase.
The symbols with error bars show UD diameters for the J\,1.24, H\,1.65, K\,2.26 and L\,3.08 filters and the dotted 
lines show the best-fitting sinusoids.
The dots show the visual magnitude from the AAVSO database averaged into ten-day bins (A.A. Henden et al. 2006, private communication), taken during the period of our observations. 
Data have been replicated for two cycles to clarify the sinusoidal pulsation.
}
\label{lc_diam_ocet_rleo}
\end{figure*}

\begin{figure*}[htbp]
\epsscale{1.7}
\plotone{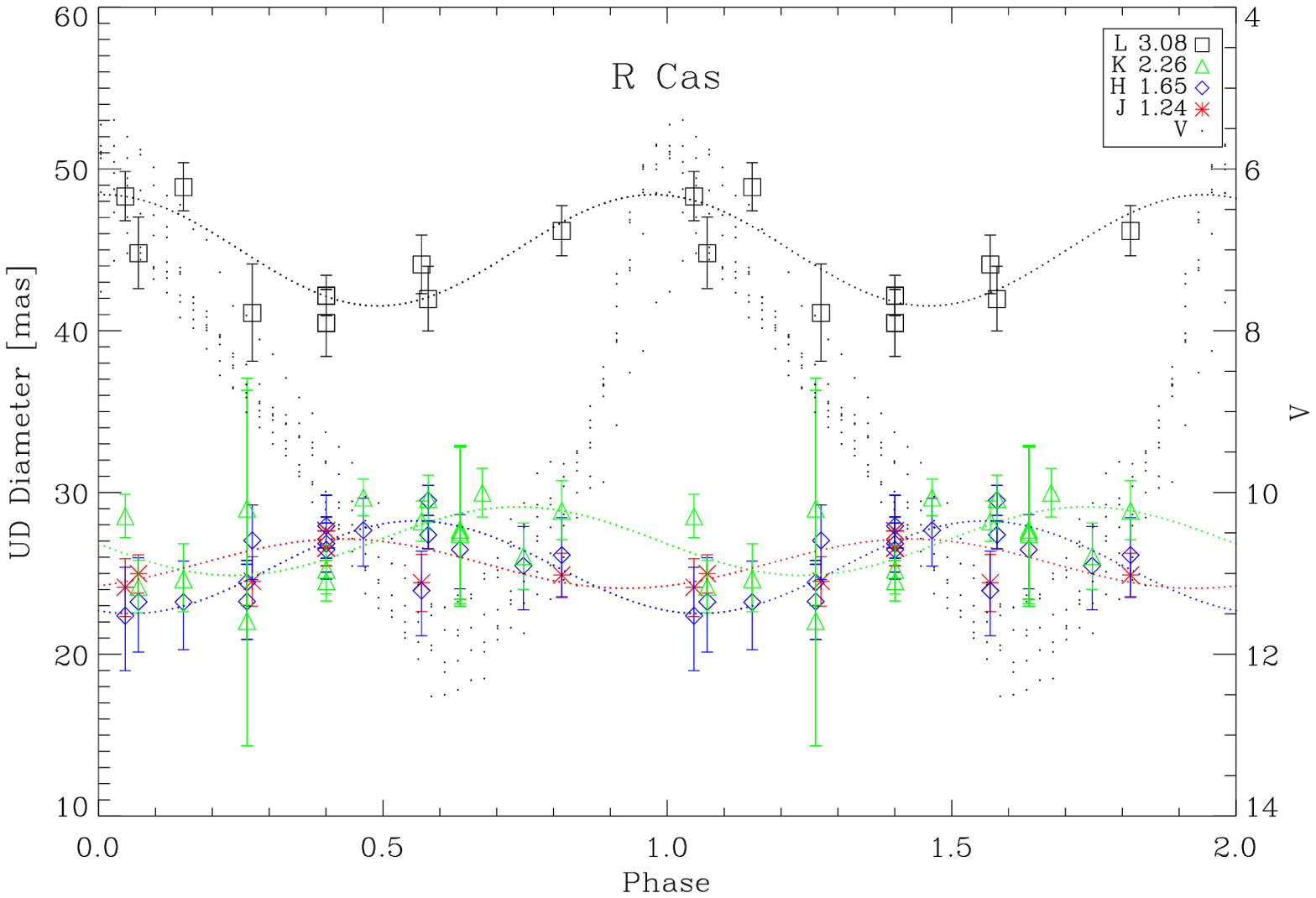}
\plotone{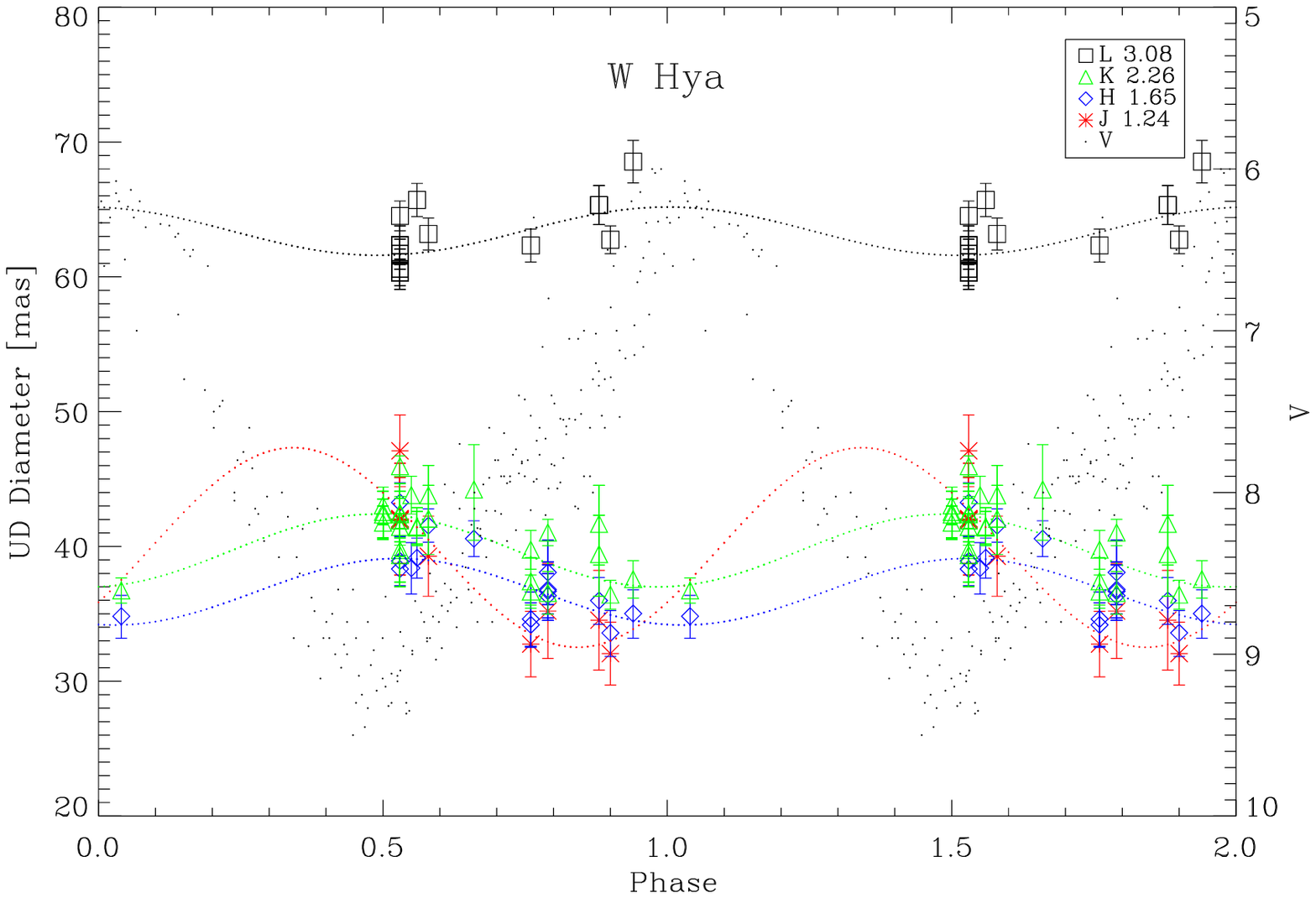}

\caption{Same as figure \ref{lc_diam_ocet_rleo} but for R Cas (upper panel) and W Hya (lower panel)}
\label{lc_diam_rcas_whya}
\end{figure*}

\begin{figure*}[htbp]
\epsscale{1.7}
\plotone{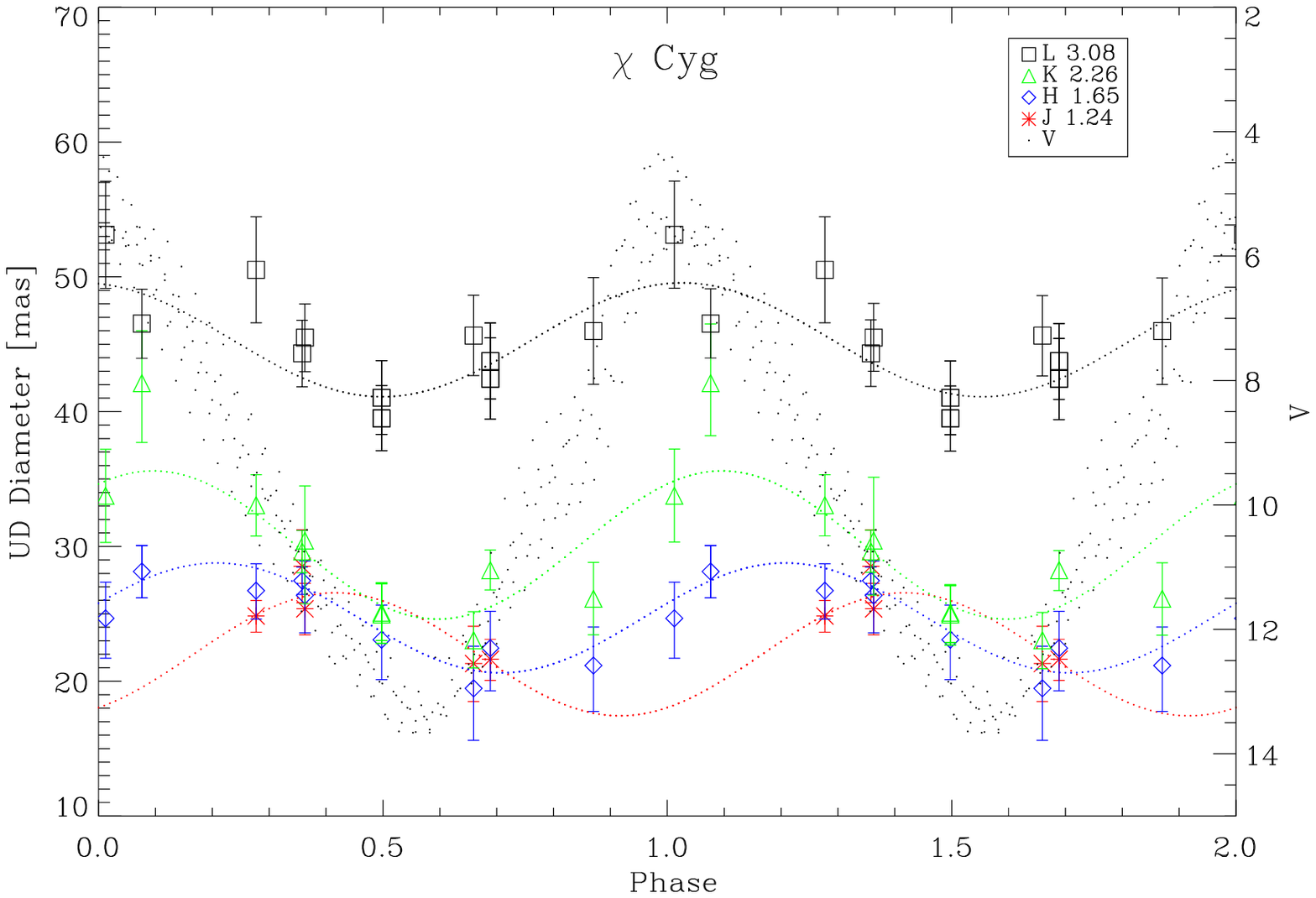}
\plotone{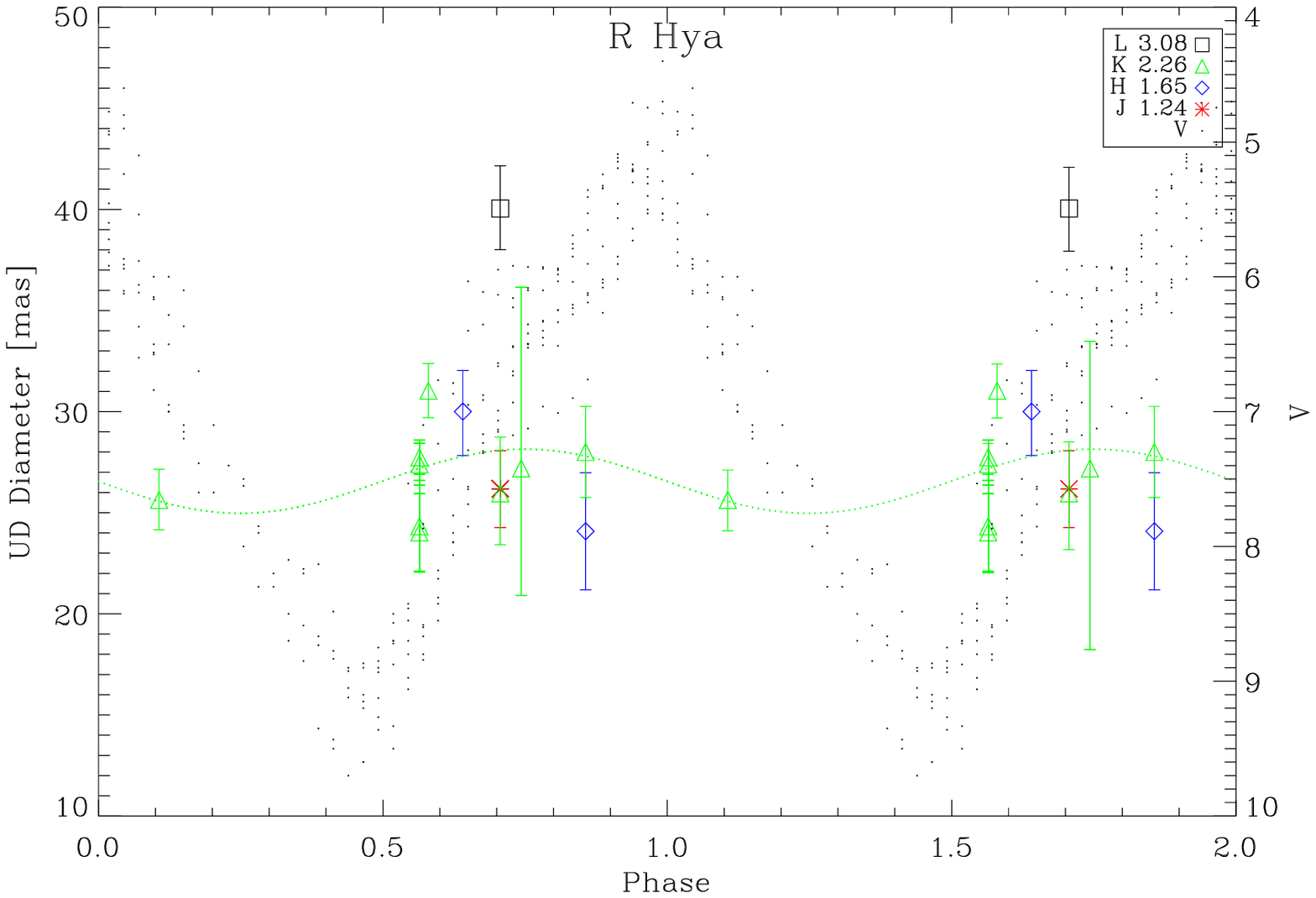}

\caption{Same as figure \ref{lc_diam_ocet_rleo} but for $\chi$ Cyg (upper panel) and R Hya (lower panel)}
\label{lc_diam_xcyg_rhya}
\end{figure*}

Figures \ref{lc_diam_ocet_rleo} to \ref{lc_diam_xcyg_rhya} show the UD angular diameters of the sample of Miras at various 
visual phases (see Tables \ref{tbl-obs_ocet} to \ref{tbl-obs_rhya}), folded with phase, together with  visual photometry data from the AAVSO (A.A. Henden et al. 2006, private communication).\\
Near-infrared pulsations of Mira stars have been observed interferometrically before, e.g. by  \cite{TUT95}; \cite{PER}; 
\cite{YOU00}; \cite{THOM}; \cite{WOO}; \cite{FED05};\cite{RAG06}, though never with the wavelength, phase and cycle coverage of this study.
We have detected diameter variations for all Miras in this study.
As discussed for the light-curves in Section \ref{lightc}, the physical diameter vs. phase curves are certainly not strictly sinusoidal.
Nonetheless, a sinusoidal shape of the curve can be observed, and although not perfect,  the sine curves fitted to the data  have been included to guide the eye.\\
With the extensive coverage of phases and cycles in four filter bandpasses,
it becomes possible to probe the stellar atmosphere both for geometric pulsation of the continuum forming layers (the so-called photosphere),
and for contamination of the continuum by molecular blanketing.
This allows us to further constrain existing theoretical models and to make more sophisticated demands on future models.
R Hya will be exempt from further discussion as the data sampling is too sparse (see Figure \ref{lc_diam_xcyg_rhya}, lower panel).\\

\epsscale{2.5}
\section{Discussion}

\begin{figure}[htbp]
\begin{center}
\epsscale{1.0}
\plotone{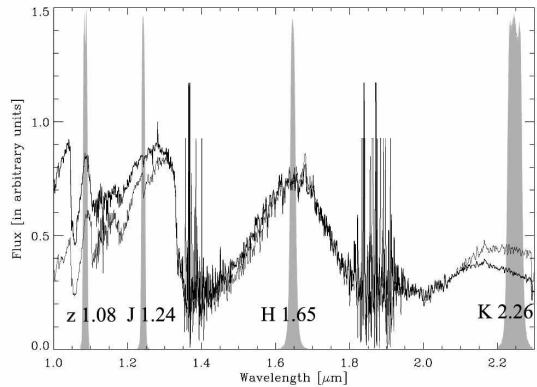}
\caption{Near-infrared spectrum of Mira R Cha at two different pulsation phases (1996May26 at phase 0.6 and 1996March3 at phase 0.3) from \cite{LW}, showing the position and shape of the z\,1.08, J\,1.24, H\,1.65 and K\,2.26 filters.
The J\,1.24 and H\,1.65 filters penetrate to layers that lie closer to the continuum forming photosphere, whereas the K\,2.26 filter sees 
a portion of the spectrum that is more contaminated by molecular opacities. }
\label{NIRspec}
\end{center}
\end{figure}
\begin{figure}[htbp]
\begin{center}
\epsscale{1.1}
\plotone{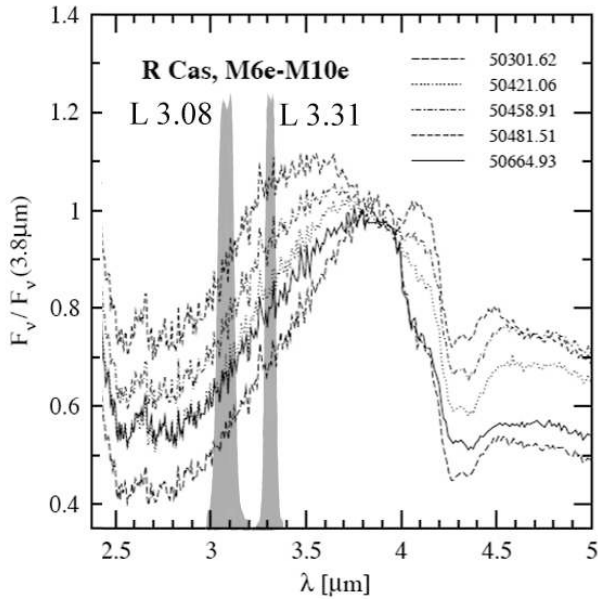}
\caption{Infrared ISO spectrum of R Cas at various phases, from \cite{ARI02} (Figure 2), with the position and shape of the L\,3.08 and L\,3.31 filters (Table \ref{filters}) overlaid. The stronger absorption,
mainly due to H$_2$O molecules, in the L\,3.08 filter comes from cooler strata (and thus further away from the continuum forming layers) and leads
to the measurement of greater angular diameters. See text (Section \ref{atmos})}
\label{Lspec}
\end{center}
\end{figure}
\begin{figure}[htbp]
\begin{center}
\epsscale{1.1}
\plotone{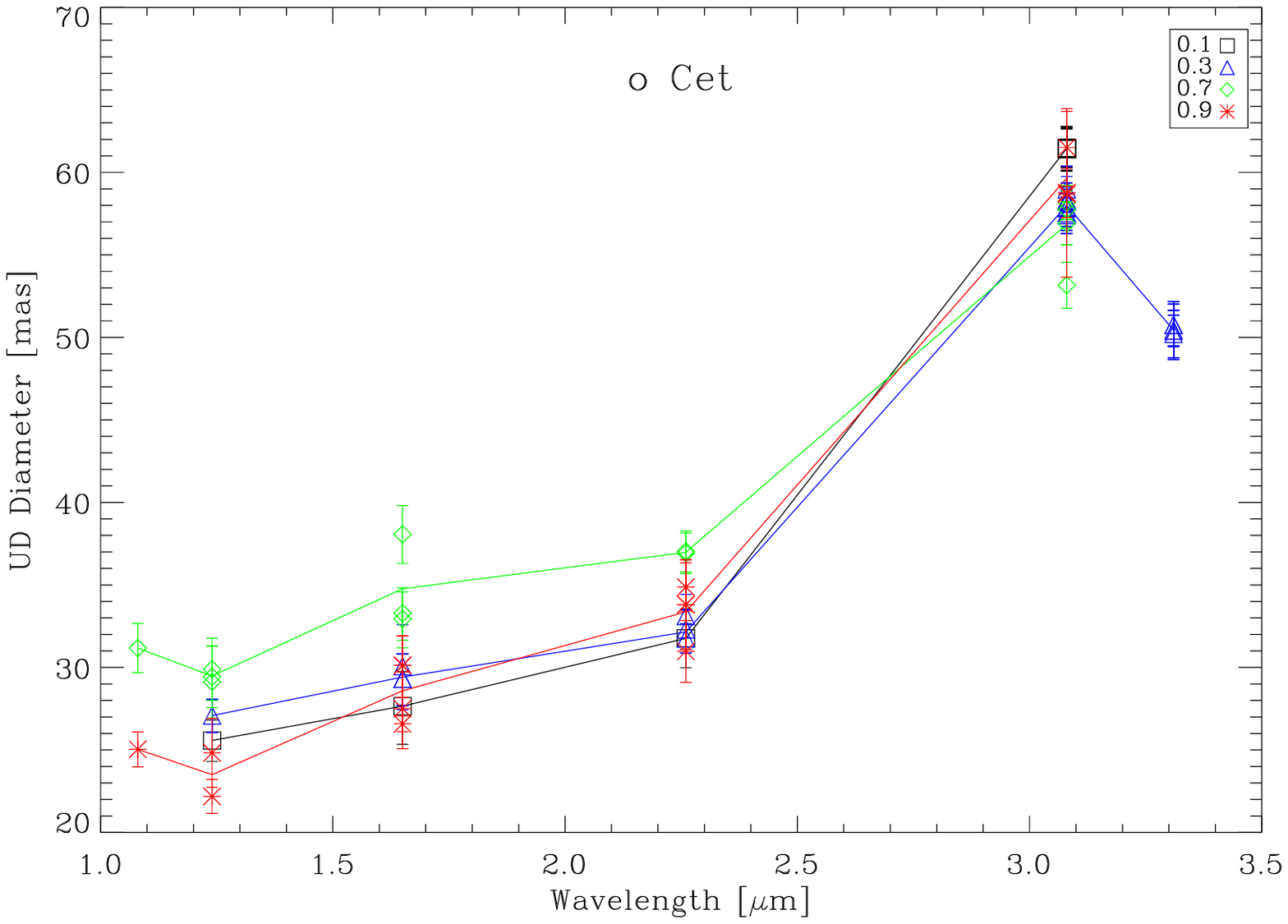}
\caption{Diameter vs. wavelength relationship for $o$~Cet measured at visual phases 0.1, 0.3, 0.7 and 0.9. All visual phase measurements carry a binning error of $\pm$ 0.1}
\label{diamp1}
\end{center}
\end{figure}

\subsection{Effects of Molecular Absorption on Multi-wavelength Diameter Observations}\label{atmos}

Figure \ref{NIRspec} shows the NIR spectrum of a Mira (R Cha) at two different pulsation phases, with the position and shape of the z\,1.08, J\,1.24, H\,1.65 and K\,2.26 filters,
and Figure \ref{Lspec} shows the  ISO spectrum of R Cas, with the position and shape of the L\,3.08 and L\,3.31 filters.
The narrow band filters J\,1.24 and H\,1.65 should show little molecular contamination by absorption bands, i.e. they are close to sampling the
continuum-forming layers (see, e.g., \citealt{TLSW}), whereas the contamination effect should be more prominent in the
 K\,2.26 and in particular in the $L$-band filters (see, e.g., \citealt{JS}; \citealt{MEN02}; \citealt{ISTW,ISW}).\\
Within the L band, our filters (L\,3.08 and L\,3.31) sample portions with different molecular absorption lines,
 originating mainly from  H$_2$O (but also from OH and SiO, \citealt{MEN02}) in Mira atmospheres.
The L\,3.08 bandpass lies deeper in the absorption feature than the L\,3.31 bandpass. 
The photons  seen through the L\,3.08 filter should therefore originate from cooler strata that lie higher in the stellar atmosphere than the L\,3.31 photons.\\

Figure \ref{diamp1} shows the diameter vs. wavelength relationship for $o$~Cet at four different phases.
It clearly shows the effect of varying absorption features on the perceived diameter.
As our filters coincidentally sample more opaque (contaminated) layers with increasing wavelength,
there is a perceived  diameter increase as a function of wavelength, with the exception of 
the J\,1.24 and L\,3.31 filters, which lie further out of the molecular absorption bands sampled by the z\,1.08 and L\,3.08 filters, respectively.
This increase of diameter with wavelength is obvious throughout the pulsation cycle for $o$~Cet, 
all other Miras showing excursions from the J\,1.24 $>$ H\,1.65 $>$ K\,2.26 diameter trend at some phases. 
In particular R Leo seems to show a much more complex layering (see Section \ref{rleo} with Figure \ref{lc_diam_ocet_rleo}).
All stars in our sample show a large ($>30$\%) increase in UD diameter between the K\,2.26 filter and the L\,3.08 filter,
first noted in the case of R Aqr by \cite{TUT00b}.
This increase indicates that the L\,3.08 filter samples a molecular layer at a considerable distance from the photosphere.
The distance between the layers varies with phase and will be further discussed for each object individually,
as it differs greatly from star to star.

\subsection{Phase dependence of Multi-wavelength Near-Infrared Diameters}\label{phase_diam}
The distinction between true photospheric pulsation, i.e.~the upward and downward motion of the continuum forming layers,
 and the effects of molecular blanketing is not trivial to unravel, even with high angular resolution data.
The varying molecular opacity is dictated by density changes as a shock front travels through the stellar atmosphere and temperature 
changes due to variations in the radiation field, causing molecules to dissociate and re-form (cf. \citealt{SCH03} and references therein).
These more or less opaque strata which vary with the pulsation phase (and between cycles) can make it difficult to derive the near-UD
 diameter of the underlying geometrically pulsating continuum layer, veiling the geometric amplitude and the phase of pulsation.
 As the brightness distributions of Miras are more complex than simple UDs, 
 changes in molecular opacities of different layers may be capable of mimicking changes in our derived diameters, 
 complicating the interpretation of our data.
 The behaviour of the S-type Mira $\chi$ Cyg differs from the other Miras in our sample in various respects and will be discussed separately.\\
 
Of the three filters within the $JHK$ bands, the atmospheric opacity reaches its highest values in the contaminated 
(albeit less contaminated than the standard $K$ bandpass) filter K\,2.26 
throughout most of the pulsation cycle, as evidenced in Figures \ref{lc_diam_ocet_rleo}-\ref{lc_diam_xcyg_rhya} by the large angular diameters.
This increased molecular opacity has been predicted by various models (e.g. \citealt{JS}; \citealt{ISTW,ISW}; \citealt{IS06}) and has been observed by, e.g., \cite{MG_MIRA_05}.\\

\begin{figure*}[htbp]
\begin{center}
\epsscale{2.2}
\plotone{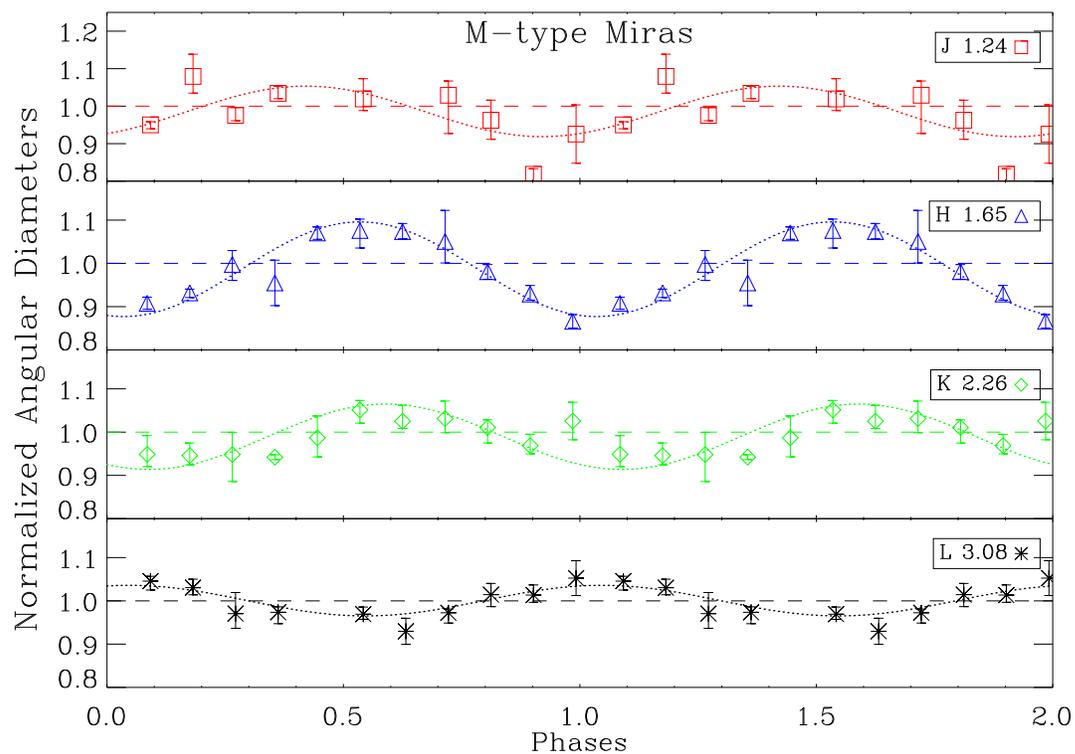}
\caption{Relative angular diameter variation as a function of phase for the 5 M-type Miras in this study.
 The sine curves represent fits to the ensemble of angular diameters divided by the mean angular diameter for each star. 
The data has been averaged into phase bins and is repeated to show two cycles for better recognition of the pulsation pattern.
The results for the different filters are offset for clarity.}
\label{adb}
\end{center}
\end{figure*}

 A diagnostic observable from our data which is more robust against sources of systematic errors (e.g. change in seeing, wind induced wobble of the telescope, calibrator characteristics)
is the relative diameter variation. 
 Figure \ref{adb} shows the relative UD diameter variation as a function of  phase for the 5 M-type Miras in this study, averaged into phase bins.
 The relative UD diameters were obtained by dividing each object's measured angular diameters by its mean,
 and the errors are representative of the scatter within each phase bin.\\
 The data illustrate the homogeneity within this sample of Miras, and allow us to investigate systematic
 differences between the observed bandpasses.
 In order to extract phase offsets and relative pulsation amplitudes in each filter, we fitted sinusoid cycloids to the combined data,
 without implying that the objects vary in such a simple fashion.\\
 
 The ensemble shows a systematic phase-shift of the relative UD diameter variation with wavelength, not unlike 
 the trend reported for the light-curves in section \ref{lightc}. 
 The J\,1.24, H\,1.65,  and K\,2.26 UD diameters reach their minimum at approximately phases 0.9, 1.0, and 1.1 respectively.
 The L\,3.08 bandpass diameters behaves differently, and will be examined in more detail at the end of this section.
 This disagrees with theoretical model predictions by \cite{ISTW, ISW}, where the diameter minimum should occur
 at roughly phases 0.7-0.8 for all bandpasses, but is comparable to the findings of \cite{THOM}, who followed the M-type
 Mira S Lac through its pulsation cycle in various sub-filters of the $K$ band.
  Using the sample as a single ``artificial'' Mira also allows us to verify the relative diameter pulsation amplitude for S Lac of \cite{THOM}, 
  who report  a 12\%-21\% peak-to-peak sinusoid amplitude in the $K$-band, with our value of 14\% peak-to-peak pulsation in the K\,2.26-band. 
  The peak-to-peak sinusoidal pulsation amplitudes for the J\,1.24, H\,1.65 and L\,3.08 bandpasses are 14\%, 22\% and 6\% respectively,
  also disagrees with theoretical models which predict much higher amplitudes (cf. \citealt{ISTW,ISW}).

 According to current models (e.g.  \citealt{ISTW,ISW}), more contaminated layers (such as the ones sampled by the K\,2.26 and L\,3.08 bandpasses)
 should experience slightly greater diameter pulsation amplitudes.\\
This trend cannot be seen in our observations: the layer experiencing the greatest diameter pulsation is the less contaminated H\,1.65 layer. 
Explanations for this could be:\\
(i) The relatively narrow width of our K\,2.26 filter compared to the standard $K$ filter used in the models,
combined with the position of our central wavelength in a possible minimum of molecular contamination as reported by \cite{THOM}.\\
(ii) A periodic change of optical depth of a layer, offset in phase to the photospheric pulsation, could result in large variations of the observed pulsation amplitude.\\
(iii) The assumed spherical symmetry could be significantly violated in the outer, more contaminated, layers (cf. \citealt{RAG06} and references therein), effectively causing departures from model predictions 
(cf., e.g., \citealt{HOF00};\citealt{IRE2004}).\\
(iv) The theoretical models need revision to accommodate for these observations.\\
 Higher molecular opacity can be expected at near-minimum phases, when the outer layers are cooler and more molecules are formed (see, e.g., \citealt{ISTW,ISW}).
For our sample stars (with the exception of $\chi$ Cyg), the maximum apparent $JHK$ UD angular diameter values are typically found near minimum visual phase,
supporting existing model interpretations.\\

\begin{figure}[htbp]
\begin{center}
\epsscale{1.1}
\plotone{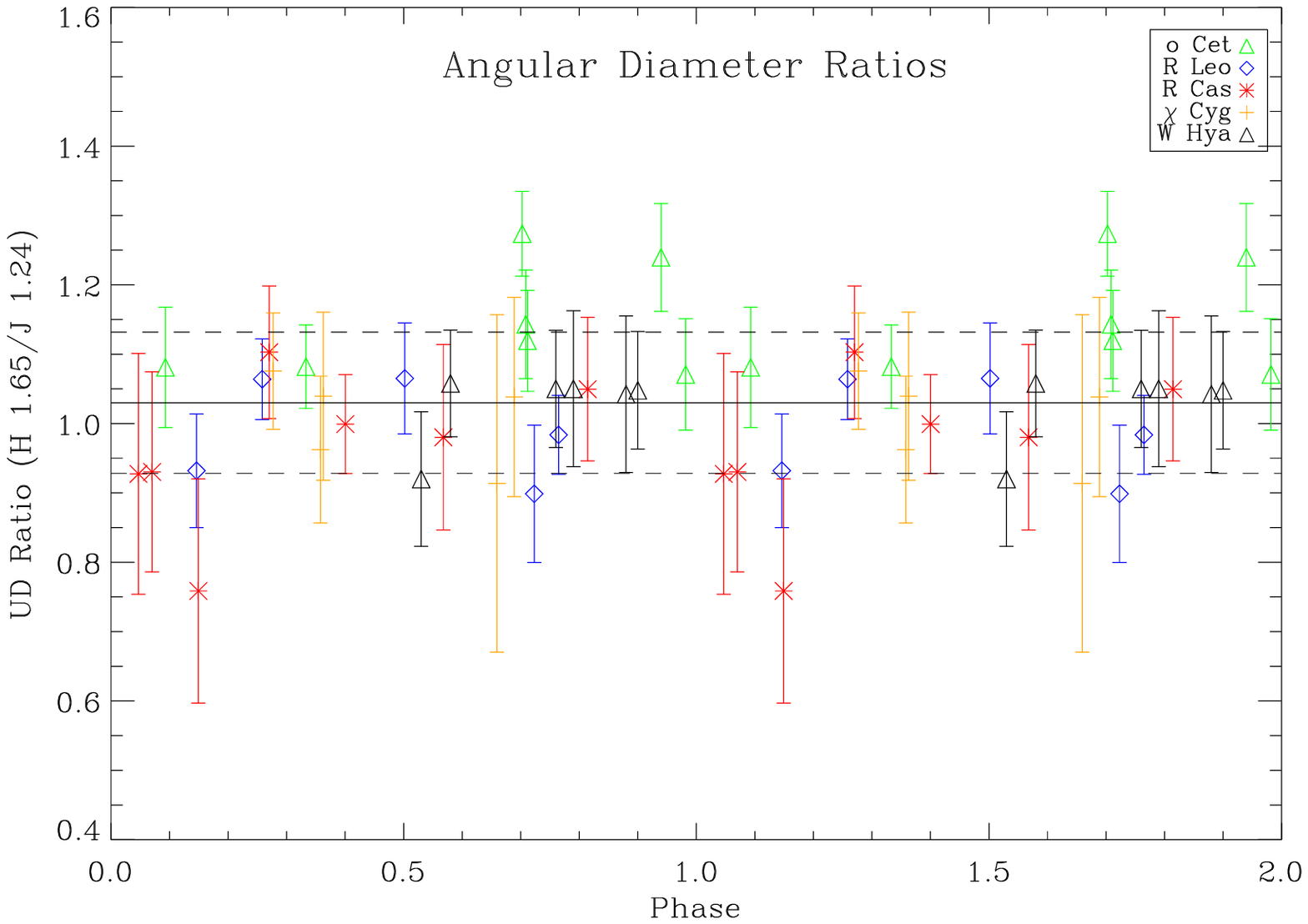}
\plotone{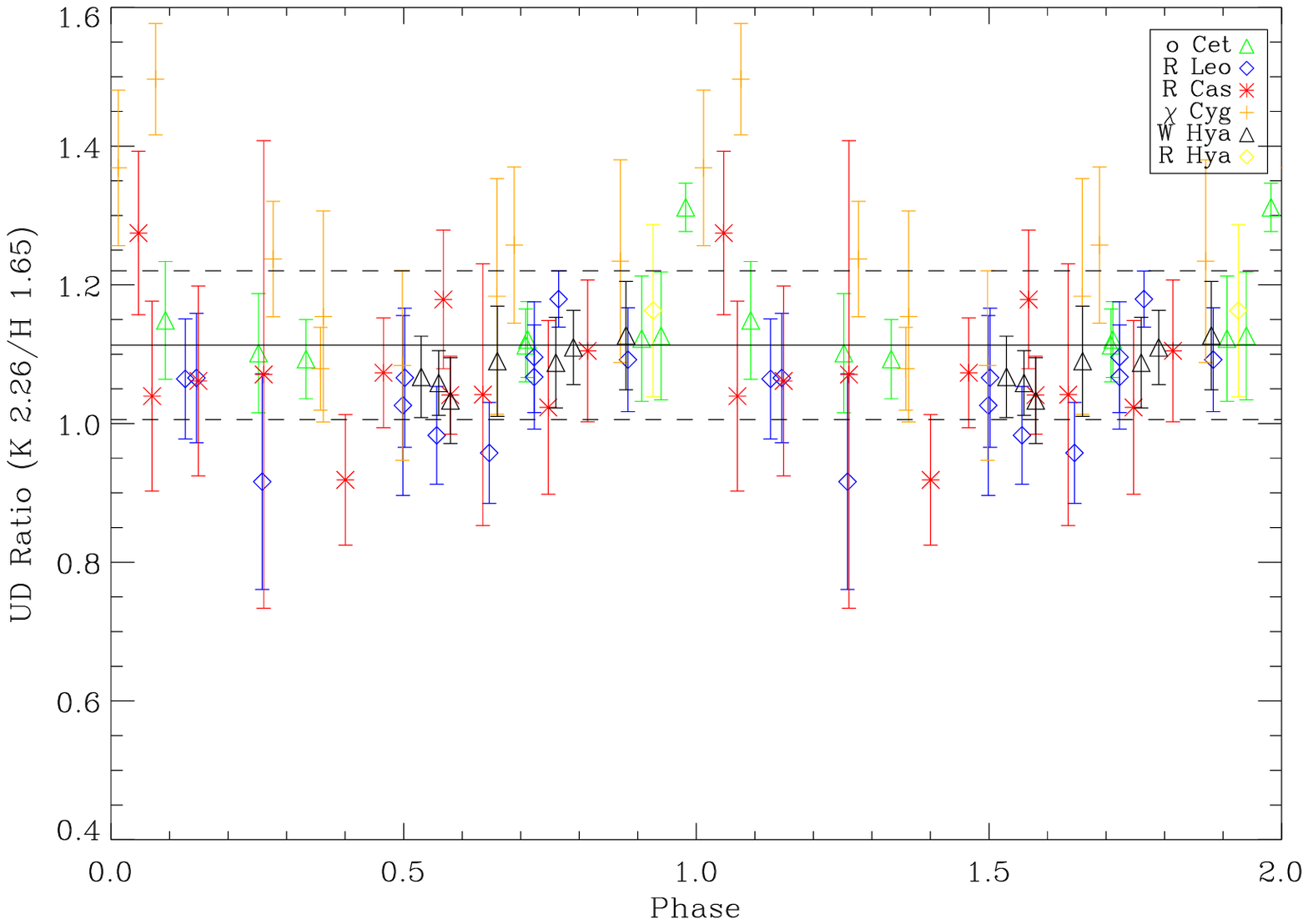}
\plotone{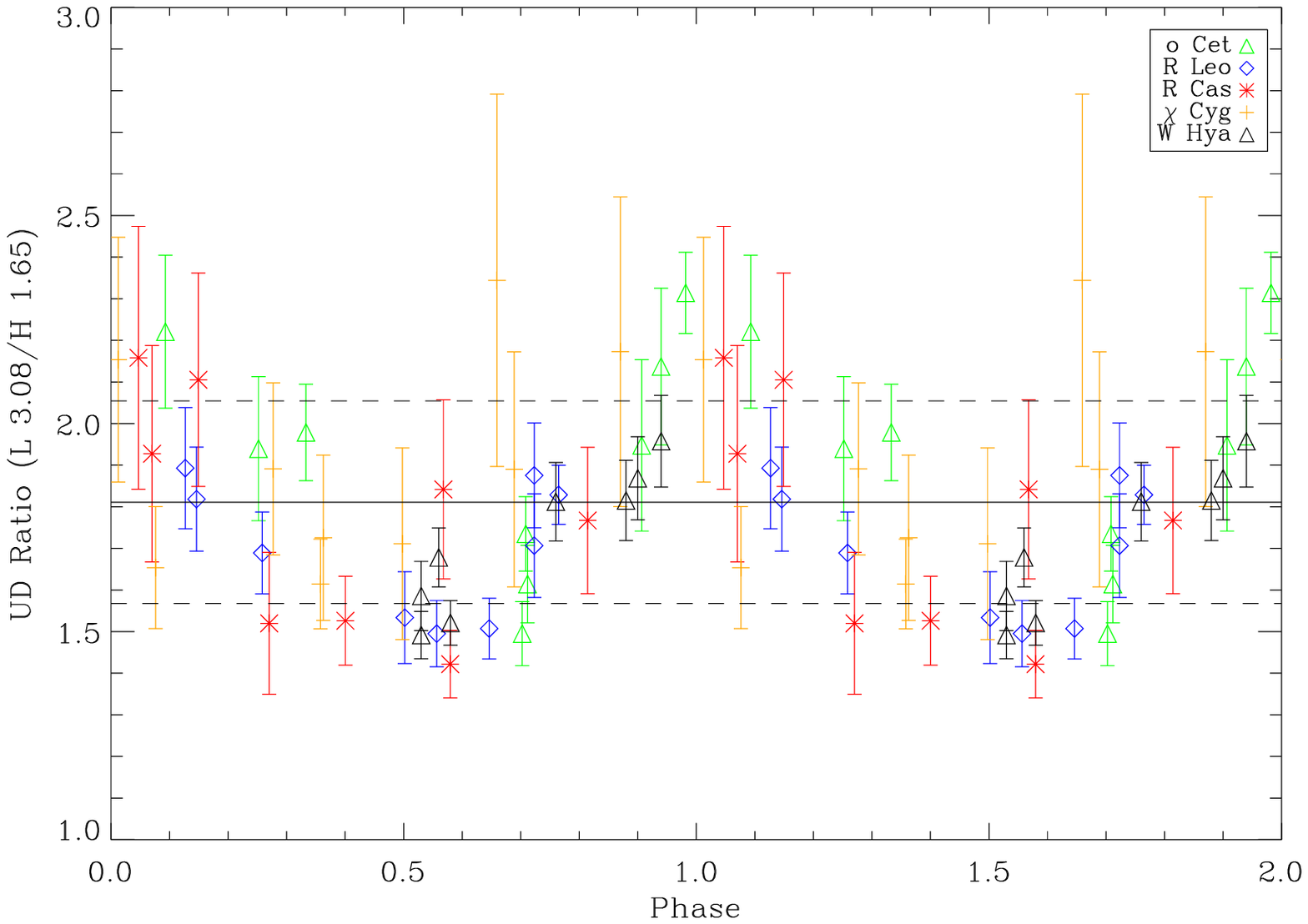}
\caption{Angular diameter ratios for all Miras in this study. The solid lines indicate the mean ratio and the dashed lines indicate the standard deviation of the sample.
The data is repeated to show two cycles for better recognition of the pulsation pattern.}
\label{fracs}
\end{center}
\end{figure}

A diagnostic observable from our data which is more robust against sources of systematic errors (e.g. change in seeing, wind induced wobble of the telescope, calibrator characteristics)
are the relative diameter ratios. 
Figure \ref{fracs} shows the UD diameter ratio  between the different filters for all our sample stars plotted vs. phase. 
There is no obvious dependence on pulsation phase of the H\,1.65/J\,1.24
quotient, though there is noteworthy scatter for our sample of 5 Miras (R Hya was never observed simultaneously in both filters) that might mask a minor phase effect .
The mean value of this diameter ratio is

\begin{equation}
\overline{R_{\rm   H\,1.65/J\,1.24}}=1.02\pm0.10
\end{equation}

\noindent which agrees, to within errors, with the value of $\overline{R_{\rm H/J }}=1.08\pm0.09$ reported by \cite{MG_MIRA_05}.
The lack of a phase dependent signature indicates a closeness in temperature and opacity variations, 
as can also be derived from the closeness in phase and pulsation amplitude seen in Figure \ref{adb}.
The diameter ratio close to unity shows the geometric closeness of the two layers.

Note that of all stars, $o$~Cet is the only one that has a ratio that includes some phase dependent effects and is slightly smaller than unity.

The ratio between the K\,2.26 and H\,1.65 filters has a more pronounced pulsation phase signature (see Figure \ref{fracs}, center panel),
which reflects more complex and disjoint temperature and opacity changes between these two layers. 
The ratio reaches its minimum (i.e. H\,1.65 UD $>$ K\,2.26 UD) before minimum light and its maximum at maximum light.
In order to compare our data with the observations of \cite{MG_MIRA_05}, we also calculated the mean diameter ratio to be

\begin{equation}
\overline{R_{\rm K\,2.26/H\,1.65}}=1.11\pm0.11,
\end{equation}

\noindent a mean ratio marginally larger that the H\,1.65/J\,1.24 mean ratio.
Again, our data agree, to within errors, with the value of  $\overline{R_{\rm K'/H }}=1.12\pm0.09$ reported by \cite{MG_MIRA_05}.\\

As mentioned earlier, the L\,3.08 angular diameter behaves differently.
It reaches minimum values at minimum light and the best fit to the diameter pulsation
 is shifted by 0.5 cycle compared with the  H\,1.65 pulsation, albeit at a much smaller relative pulsation amplitude.
As the lower panel of Figure \ref{fracs} shows, this phase shift and the UD diameter ratio between the 
two layers seems to be similar for all Miras in our sample (including $\chi$ Cyg).
For consistency sake, we calculated the mean diameter ratio to be

\begin{equation}
\overline{R_{\rm L\,3.08/H\,1.65}}=1.81\pm0.24.
\end{equation}

This unusual UD diameter variation has never been observed before and 
raises questions about the mechanism of the observed pulsation in the L\,3.08 layer.
The L\,3.08 light-curve follows a similar trend to the $JHK$ light-curves, possibly indicating a
temperature devolution similar to the lower layers, and as the inner layers of the star
are shrinking and heating up, the outer layers are either expanding or becoming increasingly opaque.
Whether opacity effects or dynamic motion of these outer layers (or both) are responsible for this surprising behaviour is the
subject for model interpretations and will be the subject of a subsequent study.

\subsection{Individual Stars}\label{indi}

Of the 6 Miras observed, 4 have observational phase coverage suitable for further comparisons with pulsation models ($o$~Cet, R Leo, R Cas and $\chi$ Cyg).
The Miras studied differ substantially in behaviour regarding pulsation amplitudes, diameter-wavelength relationships and
diameter-phase relationships

In this section we discuss the results of this study for each individual star, emphasising the main differences
and similarities found in this subset.

\subsubsubsection{$o$~Cet}\label{ocet}

The prototype of Mira stars is one of the most observed variable stars,
due to its brightness, amplitude (V$\approx10-3$, A.A. Henden et al. 2006, private communication) and closeness ($107\pm6$ pc, \citealt{KNAPP}).
Its size in different bandwidths (e.g. \citealt{HAN95};\citealt{MEN02};\citealt{WOO}), optical spectra (\citealt{JOY54}), 
lightcurves in different colors (e.g. \citealt{WHI}; \citealt{NAD01}; AAVSO), 
asymmetries (e.g. \citealt{KAR}, \citealt{TUT99}), and companion star (e.g. \citealt{KAR97};\citealt{WOO06};\citealt{IRE07}) have been subject to intense research.\\

Figure \ref{lc_diam_ocet_rleo} shows the UD diameter variation in the J\,1.24, H\,1.65 and K\,2.26 bandpasses
as nearly synchronous (within the 0.1 phase shift shown in Figure \ref{adb}), sinusoidal pulsations, with an apparent phase shift to the diameter pulsation in the L\,3.08 bandpass. 
The UD diameter vs. phase curves agree well with the fitted sine functions, with reduced $\chi^2$ for 
J\,1.24, H\,1.65, K\,2.26 and L\,3.08 having the values 1.45, 1.17, 1.02 and 0.80 respectively.
The shift of $\approx0.5$ cycles between the sine curves fitted to the diameter vs. phase in the H\,1.65 and the L\,3.08 bandpasses
has never been observed nor predicted, and can also be observed in R Leo, R Cas and W Hya.
A similar shift, though not as pronounced, can be seen in $\chi$ Cyg\\
The UD diameters vary between $22.2\pm1.0$ and $29.9\pm1.9$ mas ($\approx 35\%$) in the J\,1.24 bandpass, 
and between $53.2\pm1.4$ and $61.5\pm2.3$ mas ($\approx 16\%$) in the L\,3.08 bandbass.
The K\,2.26 UD angular diameter values lie in the range of $31.0\pm 1.9$\,mas near V-maximum and $37.0\pm1.2$\,mas at phase 0.7.
These values agree with the interferometric  UD diameters within the K-band of \cite{RIDG} (phase 0.8)
and \cite{WOO} (phases 0.1-0.4, see Table \ref{ext_diams}). 
Our UD angular diameters are generally larger than those measured by \cite{MEN02} (phases 0.9-0.0).
This is most likely due to to the larger spatial frequencies (i.e. longer baselines) at which their measurements where made,
combined with the known departure of $o$~Cet's brightness distribution from UD (cf. \citealt{WOO}). \\
 Our K\,2.26 UD diameters are also larger than the molecular layer diameters obtained by 
 \cite{PER04} (24.95$\pm0.10$ to $26.84\pm0.06$\,mas) at similar phases by fitting ad hoc
 scenarios (a photosphere surrounded by an emissive and absorbing layer) to $K$ and $L'$ interferometric data.
This is to be expected, as their visibilities also show obvious departures from simple models, and fitting the same data to a brightness distribution
consisting of a central object with a bright molecular shell would yield smaller diameters than a UD fit.
 Note that these very simple scenarios are not always unique and a new ad hoc parameter set has to be determined for each observation,
 making a comparison awkward. 
 $o$ Cet shows little cycle-to-cycle variation throughout our data, although this could be attributed to the observing of a stable
 era of its pulsation, and might have been different if observed, e.g., 10 years earlier.

\begin{deluxetable}{lccccc}
\tablewidth{0pt}
\tablecaption {\label{ext_diams}  
Published interferometric UD angular diameters in the K band for the stars in this study}
\tablehead{
 \colhead{Star} & \colhead{Reference} &  \colhead{Angular} & \colhead{Centre Wavelength } &   \colhead{Bandwidth } & \colhead{Visual} \\ 
 			&	&\colhead{Diameter [mas]}	&	\colhead{[$\mu$m]}	&\colhead{[$\mu$m]}               &\colhead{Phase}   \\        }
\startdata
$o$ Cet&\cite{RIDG}	&36.1$\pm1.4$	&    2.2		& 0.4 &0.8 \\
&\cite{MEN02}	&28.79$\pm0.10$ &2.2		&0.44&0.9\\
&			&25.73$\pm0.09$&2.03		&$\approx0.1$&0.0\\
&			&25.13$\pm0.08$&2.15		&$\approx0.1$&0.0\\
&			&25.19$\pm0.12$&2.22		&$\approx0.1$&0.0\\
&			&29.22$\pm0.12$&2.39		&$\approx0.1$&0.0\\
&			&24.40$\pm$0.11&2.16		&0.32&0.0\\
&\cite{WOO}	&29.24$\pm$0.30&2.2		&0.4&0.1\\
&			&29.53$\pm$0.30&2.2		&0.4&0.2\\
&			&30.49$\pm$0.30&2.2		&0.4&0.3\\
&			&33.27$\pm$0.33&2.2		&0.4&0.4\\
\hline
R Leo&\cite{GIA91} \tablenotemark{*}	&33.0$\pm1.3$&	2.16		&0.03&0.2\\
&\cite{PER}		&28.18$\pm0.05$&2.16		&0.32&0.2\\
&			&30.68$\pm0.05$&2.16		&0.32&0.3\\
&\cite{TEJ99} \tablenotemark{*}	&34$\pm2$	&2.36		&0.46&0.4\\
&\cite{MON04}	&30.3$\pm0.3$&2.26\&2.16	&0.05\&0.32&0.6-0.7\\
&\cite{MG_MIRA_05}&29.91$\pm0.27$&2.16	&0.32&0.4\\
&\cite{FED05}	&28.1$\pm0.05$&2.2		&0.4&0.1\\
&			&26.2$\pm$0.01&2.2		&0.4&0.0\\
\hline
R Cas&\cite{VANB02} 	&22.03$^{\mbox{\tiny+2.13}}_{\mbox{\tiny-4.14}}$&2.2&0.4& 0.8\\
\hline
W Hya&\cite{MON04}	&42.5$\pm0.7$&2.26\&2.16	&0.05\&0.32&0.5\\
&\cite{MG_MIRA_05}&39.9$\pm0.2$&2.16	&0.32&0.6\\
\hline
$\chi$ Cyg&\cite{MEN02} 	&23.24$\pm0.08$	&2.16&0.32&0.38\\
\hline
R Hya&\cite{MG_MIRA_05}&23.9$\pm0.5$&2.16	&0.32&0.8\\
&			&25.8$\pm0.2$&2.16	&0.32&0.8\\
\enddata
\tablenotetext{*}{obtained by lunar occultation}
\end{deluxetable}

  The L\,3.08 $>>$ K\,2.26 $>$ H\,1.65 $>$ J\,1.24 layering of monochrome diameters is strictly monotonous,
   unlike, e.g., R Leo, where the sequence of UD diameters seems to invert during the pulsation cycle.
 The fact that $o$~Cet appears largest in the JHK filters approximately at visual phase 0.6, roughly coinciding with the minimum of the NIR and visual light curves (see Section \ref{lightc}) is in accordance with  various model interpretations (e.g. \citealt{JS};\citealt{ISTW,ISW};\citealt{IS06}). 
 The unusual behaviour of the L\,3.08 layer, as described in Section \ref{phase_diam}, will be further discussed in Section \ref{sum} and in a follow on paper.

\subsubsubsection{R Leo}\label{rleo}

Although R Leo has a similar period (310 days), V magnitude range ($4.4-11.3$, \citealt{KHO98}) and spectral type as $o$~Cet,we found significant differences in this study.
The pulsation amplitudes of  layers in different bandpasses and their phases with respect to one another seem to show a more complex trend than $o$~Cet's atmosphere.
The pulsation is less pronounced than $o$~Cet's in the NIR, with UD diameters varying between $29.6\pm1.4$ and $33.1\pm2.0$ mas ($\approx 12\%$) in the J\,1.24 bandpass,
which shows the largest relative pulsation amplitude.
The cycle-to-cycle variation in the diameter pulsation is significant in the K\,2.26 bandpass (compared to, e.g., $o$~Cet).
Again, this effect could be due to mere coincidence, and only long term ($>20$ years) observation campaigns can ascertain this variability.
 This makes it difficult to compare our $K$ UD diameter measurements with values found in the literature (see Table \ref{ext_diams}).
There is nonetheless an overall agreement with cited UD radii except for the long baseline UD angular diameters of \cite{FED05}.
This can be due to the same effect as described in Section $o$~Cet concerning the long baseline measurements of \cite{MEN02}.
With our H\,1.65 measurements we are able to disambiguate the two possible solutions (due to ambiguous model fitting)
for the $H$-band diameter found by \cite{MG_MIRA_05}. 
Due to the seeming lack of cycle-to-cycle dependence in this bandpass, we can reject their smaller angular diameter of 23.8$\pm0.3$\,mas at phase 0.4 and
substantiate their larger diameter of 32.4$\pm0.4$\,mas\\

 The variation between $47.8\pm1.8$ and $56.5\pm2.0$\,mas ($\approx 18\%$) in the L\,3.08 bandpass shows a steady increase in diameter 
 between 97Dec16 and 01Jun11 and no significant diameter change up to 3 pulsation cycles later.
This gradual increase in angular diameter over 4 cycles may be
understandable in terms of the non-cyclic time evolution of positions of outer 
mass zones of pulsation models over several successive cycles (cf. Figures 1 of 
Ireland et al.  2004a, b), which affect the position and physics of water shells.
This interpretation is supported by the  steady brightening of the visual photometry maxima during the same 4 cycles in which the UD diameter 
increases, indicating variability in timescales longer than the pulsation phase.
The pulsation signature of the L\,3.08 bandpass shows nonetheless the general $\approx0.5$ phase lag compared to the H\,1.65 band, albeit with 
a larger uncertainty in the least squares sinusoidal fit (reduced $\chi^2=1.38$).\\

\subsubsubsection{R Cas}\label{rcas}

Of the two Miras with the longest periods in our sample, R Cas (430 days) also exhibits the latest spectral type. In fact, at minimum visual pulsation phase it defines the spectral type M10 (see \citealt{LOCK71}).
It seems to follow the model-predicted phase-diameter trend of larger diameters around minimum light
in the H\,1.65 filter, albeit with some cycle-to-cycle uncertainty. The scatter is considerably greater in the J\,1.24 and K\,2.26 filters,
but the same trend is still present.
Note that the relative phase shift of diameter maxima from shorter to longer wavelengths in $JHK$ is very distinctive,
 even when the uncertainties introduced by cycle-to-cycle variations in the diameter pulsations are taken into account. 
The position of the L\,3.08 layer also follows the pattern of the other M-type Miras, in that it seems to pulsate with
a 0.5 phase shift to the H\,1.65 layer.
\cite{VANB02} measured a $K$ UD angular diameter of 22.03$^{\mbox{\tiny+2.13}}_{\mbox{\tiny-4.14}}$\,mas at phase 0.81,  which
does not agree with our diameter of 28.9$\pm1.9$ at the same pulsation phase but agrees with the diameter measured at roughly the
same phase in a different cycle (26.1$\pm2.1$ at phase 0.74), another indicator of noticeable cycle-to-cycle variation.\\

\subsubsubsection{W Hya}\label{whya}
At a distance of 78$\pm$3 pc (\citealt{KNAPP}), W Hya is the closet and best resolved Mira in our sample. 
We observe  a relatively time independent K\,2.26 angular diameter of $\approx40\pm5$\,mas in the phase range $\Phi=0.53-0.88$.
Because of W Hya's low declination and proximity to the ecliptic, and because the period is close to 1 year, 
the phase coverage only spans half the pulsation cycle (see Table \ref{tbl-obs_whya}).
In this short time interval, covering only 0.44 phases, the J\,1.24 diameter shrinks from 47.1$\pm2.7$\,mas to $32.0\pm2.3$\,mas, a 32\% decrease, whereas the 
H\,1.65 diameter decreases by 22\% and the K\,2.26 diameter shows a decrease of $\approx$20\%.\\
The previously measured UD angular diameters in the $K$-band by \cite{MON04} and \cite{MG_MIRA_05} (see Table \ref{ext_diams}) are consistent
with our measurements.  The $H$-band diameter of \cite{MG_MIRA_05} (31.3$\pm0.3$\,mas) is slightly smaller than our derived H\,1.65 UD angular diameters,
which is most likely due to the same effects of long baseline interferometry as described for $o$~Cet and R Leo.
Although all Mira intensity distributions show deviations from a UD profile, we can clearly detect those only in W Hya. 
This could be due to the presence of dust emission and/or the partially resolved brightness distribution of
extended molecular layers in the upper atmosphere.

The largest L\,3.08 angular diameter is found around maximum visual light,
where the J\,1.24 and H\,1.65 diameters are smallest.

\subsubsubsection{$\chi$ Cyg}\label{xcyg}
The Mira in our sample with the second longest pulsation period, $\chi$ Cyg (408 days) is the only S-type Mira
,albeit with M-type characteristics (cf. \citealt{KEE80}).
It deviates noticeably from the M-type Miras in many ways.
The visual light-curve of $\chi$ Cyg is well known for its particularly large magnitude range, and
its near-IR colours are also all unusually red, as observed by \cite{WHI}.
The L\,3.08 mean magnitude, in particular, is comparable to \textit{or brighter than} that of K\,2.26,
whereas for all other Miras in our study the L\,3.08 magnitudes lie in the rage between the H\,1.65 and the K\,2.26 light-curves.\\

The relative diameter pulsation amplitude is larger in all filters than that of the other Miras in this paper, and the 
derived UD angular diameters show small cycle-to-cycle variation.
There are no model predictions for S-type Miras to date, and we can only speculate whether or not
this behaviour is related to the fundamental stellar parameters or to the fact that differing C to O ratio lead to 
significant changes in H$_2$O formation characteristics (cf. \citealt{OHN04}) and more stable opacity structures.

While the L\,3.08 UD diameter pulsation follows the same trend as $o$~Cet, R Leo, R Cas and W Hya,
in that it reaches maximum amplitude around maximum visual light, the $JHK$ diameters show a somewhat different behaviour.
Note that the shapes of the visibility curves deviate increasingly from a UD profile with increasing wavelength.
The K\,2.26 and H\,1.65 UD angular diameters reach their maximum values just after maximum light, at phase 0.1-0.2,
coinciding with the maximum of the correspondent light curves. This may indicate a heavy contamination of these
filters by molecular layers in the star's atmosphere.
On the other hand, the J\,1.24 band data infers a smaller UD diameter with its maximum around minimum light, as observed in
the other Miras in our sample, suggesting a deeper view into $\chi$ Cyg's atmosphere.

\cite{YOU00} only detected a slight variation in the $J$-band angular diameters,
albeit over a very small phase coverage.
If we fit a sine curve to our derived J\,1.24 angular diameters, then we find that \cite{YOU00} measurements lie roughly symmetrically around the 
diameter pulsation minimum, a factor that could explain the lack of diameter variation in their studies.
We converted the $J$-band Gaussian FWHM from \cite{YOU00} to UD angular diameters using the UD to Gaussian ratio
$R\approx1.5$ calculated by \cite{BURNS}, and find that the diameters approximately
match our simple sine curve predictions. 
\cite{MEN02} report a $K'$ (centered at 2.16\,$\mu$m, 0.32$\mu$m wide) UD angular diameter of 23.24$\pm0.08$\,mas for $\chi$ Cyg on 2000 May, at variable phase 0.38.
On June of the same year (phase 0.48) we obtained a K\,2.26 diameter of 25.0$\pm2.2$, in reasonable agreement.
In the next two subsequent years (both at phase 0.36) we observed UD diameters of approximately 30\,mas,
an increase in diameter of 20\% over one cycle, which might be linked to transient opacity structures.

\section{Summary}\label{sum}

We have measured the diameters and NIR light-curves  of 6 Miras stars at up to 19 separate phases in 4 filters, the first study of this magnitude.
We present the first narrowband 3.08\,$\mu$m light-curves of Miras.
The NIR light-curves can be approximated by a sine function and confirm the phase shift of $\approx 0.15 - 0.22$ by which the NIR maxima
lag behind the visual maxima previously reported by \cite{NAD01} and \cite{SMITH02}.
In addition we find a NIR photometric flux relation: J\,1.24 $<$ H\,1.65 $<$ K\,2.26, in
agreement with the $JHK$ observations of \cite{WHI} and \cite{SMITH02}.
The L\,3.08 fluxes in our sample are slightly less or equal to
the K\,2.26 fluxes, except for $\chi$ Cyg, where the L\,3.08 magnitudes are of the comparable to the ones
in the K\,2.26 bandpass.

We found no correlation between NIR photometry and UD diameter cycle-to-cycle variations,
yet found some correlation between the L\,3.08 UD diameters and the visual light-curves (see section R Leo).\\
All observed stars show variations of their UD angular diameters as a function of pulsation phase.
We find the UD diameter relation J\,1.24 $<$ H\,1.65 $<$ K\,2.26 to be an average value only, 
with deviations throughout the pulsation cycle, revealing the complexity of phase dependant opacity contamination from molecules in different layers. 
Of the Miras in our sample, only $o$~Cet shows this strict layering throughout its cycle, in agreement with theoretical models
designed to represent $o$~Cet (see \citealt{JS}).\\
The pulsation amplitude also does not follow the model-predicted dependency on molecular opacity (\citealt{ISTW,ISW}).
The layers exhibiting the largest relative variation in UD diameter (6\%-18\%) were those seen through the H\,1.65 and J\,1.24 filters, 
which should display less molecular contamination and thus less diameter variation.
The K\,2.26 layer has UD diameters pulsation amplitudes between 4\% and 7\%.\\
The molecular layer probed by the L\,3.08 bandpass is significantly further from the photosphere than the $JHK$ layers.
It varies in UD angular diameter by as little a 4\% and as much as 8\%, with a 0.5 phase offset to the H\,1.65 pulsation,
and is between 1.5 and 2.5 times larger than the H\,1.65 UD angular diameter, depending on the pulsation phase.
This behaviour has not been observed before, and detailed models are needed to understand it.\\

Another quantity predicted by models are the diameters in different wavelengths at different phases.
When comparing the two stars whose parameters are thought to resemble the model input parameters the most,
R Leo and $o$ Cet (both with revised Hipparcos parallaxes, see \citealt{KNAPP}), with model predictions (\citealt{ISTW,ISW}),
we find that our UD diameters are significantly too large in the K\,2.26 bandpass. 
This could be explained by too high model effective temperatures, but a more careful, model based interpretation is needed
to understand this effect.\\
Given the small baselines ($<$ 10\,m) used in this experiment, it is remarkable how consistent the data is for the M-type Miras.
It is even more surprising how different the S-type Mira $\chi$ Cyg appears to be when it's mutli-wavelength pulsation signature is analyzed.
The H\,1.65 and K\,2.26 UD angular diameters are smallest around minimum light, contrary to all models and previous observations of M-type Miras.
Further work with these data, including model comparisons with individual stars, imaging and asymmetry studies, will be presented in subsequent publications.

\acknowledgments

This work has been supported by grants from the National Science Foundation 
, the Australian Research Council and the Deutsche
Forschungsgemeinschaft (HCW, MS). The data presented herein were
obtained at the W.M. Keck Observatory, which is operated as a
scientific partnership among the California Institute of Technology,
the University of California and the National Aeronautics and Space
Administration.  The Observatory was made possible by the generous
financial support of the W.M. Keck Foundation.
We acknowledge with thanks the variable star observations from the AAVSO International Database contributed by observers worldwide and used in this research.
We also thank Albert Jones and Peter Williams for the W Hya light curve data.

\bibliographystyle{apj}
\bibliography{REFPAPER}

\begin{thebibliography}{68}
\expandafter\ifx\csname natexlab\endcsname\relax\def\natexlab#1{#1}\fi

\bibitem[{{Aringer} {et~al.}(2002){Aringer}, {Kerschbaum}, \&
  {J{\"o}rgensen}}]{ARI02}
{Aringer}, B., {Kerschbaum}, F., \& {J{\"o}rgensen}, U.~G. 2002, \aap, 395, 915

\bibitem[{{Bessell} {et~al.}(1996){Bessell}, {Scholz}, \& {Wood}}]{BSW}
{Bessell}, M.~S., {Scholz}, M., \& {Wood}, P.~R. 1996, \aap, 307, 481

\bibitem[{{Burns} {et~al.}(1998){Burns}, {Baldwin}, {Boysen}, {Haniff},
  {Lawson}, {Mackay}, {Rogers}, {Scott}, {St.-Jacques}, {Warner}, {Wilson}, \&
  {Young}}]{BURNS}
{Burns}, D., {Baldwin}, J.~E., {Boysen}, R.~C., {Haniff}, C.~A., {Lawson},
  P.~R., {Mackay}, C.~D., {Rogers}, J., {Scott}, T.~R., {St.-Jacques}, D.,
  {Warner}, P.~J., {Wilson}, D.~M.~A., \& {Young}, J.~S. 1998, \mnras, 297, 462

\bibitem[{{Danchi} {et~al.}(1994){Danchi}, {Bester}, {Degiacomi}, {Greenhill},
  \& {Townes}}]{DAN94}
{Danchi}, W.~C., {Bester}, M., {Degiacomi}, C.~G., {Greenhill}, L.~J., \&
  {Townes}, C.~H. 1994, \aj, 107, 1469

\bibitem[{{di Giacomo} {et~al.}(1991){di Giacomo}, {Lisi}, {Calamai}, \&
  {Richichi}}]{GIA91}
{di Giacomo}, A., {Lisi}, F., {Calamai}, G., \& {Richichi}, A. 1991, \aap, 249,
  397

\bibitem[{{Dumm} \& {Schild}(1998)}]{DUMM}
{Dumm}, T. \& {Schild}, H. 1998, New Astronomy, 3, 137

\bibitem[{{Dyck} {et~al.}(1996){Dyck}, {Benson}, {van Belle}, \&
  {Ridgway}}]{DYCK96}
{Dyck}, H.~M., {Benson}, J.~A., {van Belle}, G.~T., \& {Ridgway}, S.~T. 1996,
  \aj, 111, 1705

\bibitem[{{Dyck} {et~al.}(1998){Dyck}, {van Belle}, \& {Thompson}}]{DYCK98}
{Dyck}, H.~M., {van Belle}, G.~T., \& {Thompson}, R.~R. 1998, \aj, 116, 981

\bibitem[{{Fedele} {et~al.}(2005){Fedele}, {Wittkowski}, {Paresce}, {Scholz},
  {Wood}, \& {Ciroi}}]{FED05}
{Fedele}, D., {Wittkowski}, M., {Paresce}, F., {Scholz}, M., {Wood}, P.~R., \&
  {Ciroi}, S. 2005, \aap, 431, 1019

\bibitem[{{Golay}(1971)}]{GOL71}
{Golay}, M. J.~E. 1971, J. Opt. Soc. Am., 61, 272

\bibitem[{{Hanbury Brown} {et~al.}(1974){Hanbury Brown}, {Davis}, \&
  {Allen}}]{HAN74}
{Hanbury Brown}, R., {Davis}, J., \& {Allen}, L.~R. 1974, \mnras, 167, 121

\bibitem[{{Haniff} {et~al.}(1995){Haniff}, {Scholz}, \& {Tuthill}}]{HAN95}
{Haniff}, C.~A., {Scholz}, M., \& {Tuthill}, P.~G. 1995, \mnras, 276, 640

\bibitem[{{Hofmann} {et~al.}(2000){Hofmann}, {Beckmann}, {Bl{\" o}cker},
  {Coud{\' e} du Foresto}, {Lacasse}, {Millan-Gabet}, {Morel}, {Pras},
  {Ruilier}, {Schertl}, {Scholz}, {Shenavrin}, {Traub}, {Weigelt},
  {Wittkowski}, \& {Yudin}}]{HOF00}
{Hofmann}, K.-H., {Beckmann}, U., {Bl{\" o}cker}, T., {Coud{\' e} du Foresto},
  V., {Lacasse}, M.~G., {Millan-Gabet}, R., {Morel}, S., {Pras}, B., {Ruilier},
  C., {Schertl}, D., {Scholz}, M., {Shenavrin}, V., {Traub}, W.~A., {Weigelt},
  G., {Wittkowski}, M., \& {Yudin}, B. 2000, in Interferometry in Optical
  Astronomy, eds.~P.~J.~Lena and A.~Quirrenbach, SPIE Proc., 4006, 688

\bibitem[{{Hofmann} {et~al.}(2002){Hofmann}, {Beckmann}, {Bl{\"o}cker},
  {Coud{\'e} du Foresto}, {Lacasse}, {Mennesson}, {Millan-Gabet}, {Morel},
  {Perrin}, {Pras}, {Ruilier}, {Schertl}, {Sch{\"o}ller}, {Scholz},
  {Shenavrin}, {Traub}, {Weigelt}, {Wittkowski}, \& {Yudin}}]{HOF02}
{Hofmann}, K.-H., {Beckmann}, U., {Bl{\"o}cker}, T., {Coud{\'e} du Foresto},
  V., {Lacasse}, M., {Mennesson}, B., {Millan-Gabet}, R., {Morel}, S.,
  {Perrin}, G., {Pras}, B., {Ruilier}, C., {Schertl}, D., {Sch{\"o}ller}, M.,
  {Scholz}, M., {Shenavrin}, V., {Traub}, W., {Weigelt}, G., {Wittkowski}, M.,
  \& {Yudin}, B. 2002, New Astronomy, 7, 9

\bibitem[{{Hofmann} {et~al.}(1998){Hofmann}, {Scholz}, \& {Wood}}]{HSW}
{Hofmann}, K.-H., {Scholz}, M., \& {Wood}, P.~R. 1998, \aap, 339, 846

\bibitem[{{Ireland}(2006)}]{IRE2006}
{Ireland}, M.~J. 2006, in Advances in Stellar Interferometry. Edited by
  Monnier, John D.; Sch{\"o}ller, Markus; Danchi, William C.. Proceedings of
  the SPIE, Volume 6268, pp. (2006).

\bibitem[{{Ireland} {et~al.}(2007){Ireland}, {Monnier}, {Tuthill}, {Cohen}, {De
  Buizer}, {Packham}, {Ciardi}, {Hayward}, \& {Lloyd}}]{IRE07}
{Ireland}, M.~J., {Monnier}, J.~D., {Tuthill}, P.~G., {Cohen}, R.~W., {De
  Buizer}, J.~M., {Packham}, C., {Ciardi}, D., {Hayward}, T., \& {Lloyd}, J.~P.
  2007, ArXiv Astrophysics e-prints

\bibitem[{{Ireland} \& {Scholz}(2006)}]{IS06}
{Ireland}, M.~J. \& {Scholz}, M. 2006, \mnras, 367, 1585

\bibitem[{{Ireland} {et~al.}(2004{\natexlab{a}}){Ireland}, {Scholz}, {Tuthill},
  \& {Wood}}]{ISTW}
{Ireland}, M.~J., {Scholz}, M., {Tuthill}, P.~G., \& {Wood}, P.~R.
  2004{\natexlab{a}}, \mnras, 355, 444

\bibitem[{{Ireland} {et~al.}(2004{\natexlab{b}}){Ireland}, {Scholz}, \&
  {Wood}}]{ISW}
{Ireland}, M.~J., {Scholz}, M., \& {Wood}, P.~R. 2004{\natexlab{b}}, \mnras,
  352, 318

\bibitem[{{Ireland} {et~al.}(2004{\natexlab{c}}){Ireland}, {Tuthill},
  {Bedding}, {Robertson}, \& {Jacob}}]{IRE2004}
{Ireland}, M.~J., {Tuthill}, P.~G., {Bedding}, T.~R., {Robertson}, J.~G., \&
  {Jacob}, A.~P. 2004{\natexlab{c}}, \mnras, 350, 365

\bibitem[{{Jacob} \& {Scholz}(2002)}]{JS}
{Jacob}, A.~P. \& {Scholz}, M. 2002, \mnras, 336, 1377

\bibitem[{{Joy}(1954)}]{JOY54}
{Joy}, A.~H. 1954, \apjs, 1, 39

\bibitem[{{Jura} \& {Kleinmann}(1990)}]{JK90}
{Jura}, M. \& {Kleinmann}, S.~G. 1990, \apjs, 73, 769

\bibitem[{{Karovska} {et~al.}(1997){Karovska}, {Hack}, {Raymond}, \&
  {Guinan}}]{KAR97}
{Karovska}, M., {Hack}, W., {Raymond}, J., \& {Guinan}, E. 1997, \apjl, 482,
  L175

\bibitem[{{Karovska} {et~al.}(1991){Karovska}, {Nisenson}, {Papaliolios}, \&
  {Boyle}}]{KAR}
{Karovska}, M., {Nisenson}, P., {Papaliolios}, C., \& {Boyle}, R.~P. 1991,
  \apjl, 374, L51

\bibitem[{{Keenan} \& {Boeshaar}(1980)}]{KEE80}
{Keenan}, P.~C. \& {Boeshaar}, P.~C. 1980, \apjs, 43, 379

\bibitem[{{Kholopov} {et~al.}(1998){Kholopov}, {Samus}, {Frolov}, {Goranskij},
  {Gorynya}, {Karitskaya}, {Kazarovets}, {Kireeva}, {Kukarkina}, {Kurochkin},
  {Medvedeva}, {Pastukhova}, {Perova}, {Rastorguev}, \& {Shugarov}}]{KHO98}
{Kholopov}, P.~N., {Samus}, N.~N., {Frolov}, M.~S., {Goranskij}, V.~P.,
  {Gorynya}, N.~A., {Karitskaya}, E.~A., {Kazarovets}, E.~V., {Kireeva}, N.~N.,
  {Kukarkina}, N.~P., {Kurochkin}, N.~E., {Medvedeva}, G.~I., {Pastukhova},
  E.~N., {Perova}, N.~B., {Rastorguev}, A.~S., \& {Shugarov}, S.~Y. 1998, in
  Combined General Catalogue of Variable Stars, 4.1 Ed (II/214A). (1998)

\bibitem[{{Knapp} {et~al.}(2003){Knapp}, {Pourbaix}, {Platais}, \&
  {Jorissen}}]{KNAPP}
{Knapp}, G.~R., {Pourbaix}, D., {Platais}, I., \& {Jorissen}, A. 2003, \aap,
  403, 993

\bibitem[{{Lan{\c c}on} \& {Wood}(2000)}]{LW}
{Lan{\c c}on}, A. \& {Wood}, P.~R. 2000, \aaps, 146, 217

\bibitem[{{Lockwood} \& {Wing}(1971)}]{LOCK71}
{Lockwood}, G.~W. \& {Wing}, R.~F. 1971, \apj, 169, 63

\bibitem[{{Mennesson} {et~al.}(2002){Mennesson}, {Perrin}, {Chagnon}, {du
  Coud{\'e} Foresto}, {Ridgway}, {Merand}, {Salome}, {Borde}, {Cotton},
  {Morel}, {Kervella}, {Traub}, \& {Lacasse}}]{MEN02}
{Mennesson}, B., {Perrin}, G., {Chagnon}, G., {du Coud{\'e} Foresto}, V.,
  {Ridgway}, S., {Merand}, A., {Salome}, P., {Borde}, P., {Cotton}, W.,
  {Morel}, S., {Kervella}, P., {Traub}, W., \& {Lacasse}, M. 2002, \apj, 579,
  446

\bibitem[{{Millan-Gabet} {et~al.}(2005){Millan-Gabet}, {Pedretti}, {Monnier},
  {Schloerb}, {Traub}, {Carleton}, {Lacasse}, \& {Segransan}}]{MG_MIRA_05}
{Millan-Gabet}, R., {Pedretti}, E., {Monnier}, J.~D., {Schloerb}, F.~P.,
  {Traub}, W.~A., {Carleton}, N.~P., {Lacasse}, M.~G., \& {Segransan}, D. 2005,
  \apj, 620, 961

\bibitem[{{Monnier}(1999)}]{MON99t}
{Monnier}, J.~D. 1999, ph.D. thesis, University of California, Berkeley

\bibitem[{{Monnier} {et~al.}(2004){Monnier}, {Millan-Gabet}, {Tuthill},
  {Traub}, {Carleton}, {Coud{\'e} du Foresto}, {Danchi}, {Lacasse}, {Morel},
  {Perrin}, {Porro}, {Schloerb}, \& {Townes}}]{MON04}
{Monnier}, J.~D., {Millan-Gabet}, R., {Tuthill}, P.~G., {Traub}, W.~A.,
  {Carleton}, N.~P., {Coud{\'e} du Foresto}, V., {Danchi}, W.~C., {Lacasse},
  M.~G., {Morel}, S., {Perrin}, G., {Porro}, I.~L., {Schloerb}, F.~P., \&
  {Townes}, C.~H. 2004, \apj, 605, 436

\bibitem[{{Monnier} {et~al.}(2002){Monnier}, {Tuthill}, \& {Danchi}}]{MON02}
{Monnier}, J.~D., {Tuthill}, P.~G., \& {Danchi}, W.~C. 2002, \apjl, 567, L137

\bibitem[{{Mozurkewich} {et~al.}(2003){Mozurkewich}, {Armstrong}, {Hindsley},
  {Quirrenbach}, {Hummel}, {Hutter}, {Johnston}, {Hajian}, {Elias}, {Buscher},
  \& {Simon}}]{MOZ03}
{Mozurkewich}, D., {Armstrong}, J.~T., {Hindsley}, R.~B., {Quirrenbach}, A.,
  {Hummel}, C.~A., {Hutter}, D.~J., {Johnston}, K.~J., {Hajian}, A.~R.,
  {Elias}, II, N.~M., {Buscher}, D.~F., \& {Simon}, R.~S. 2003, \aj, 126, 2502

\bibitem[{{Mozurkewich} {et~al.}(1991){Mozurkewich}, {Johnston}, {Simon},
  {Bowers}, {Gaume}, {Hutter}, {Colavita}, {Shao}, \& {Pan}}]{MOZ91}
{Mozurkewich}, D., {Johnston}, K.~J., {Simon}, R.~S., {Bowers}, P.~F., {Gaume},
  R., {Hutter}, D.~J., {Colavita}, M.~M., {Shao}, M., \& {Pan}, X.~P. 1991,
  \aj, 101, 2207

\bibitem[{{Nadzhip} {et~al.}(2001){Nadzhip}, {Tatarnikov}, {Shenavrin},
  {Weigelt}, \& {Yudin}}]{NAD01}
{Nadzhip}, A.~E., {Tatarnikov}, A.~M., {Shenavrin}, V.~I., {Weigelt}, G., \&
  {Yudin}, B.~F. 2001, Astronomy Letters, 27, 324

\bibitem[{{Ohnaka}(2004)}]{OHN04}
{Ohnaka}, K. 2004, \aap, 424, 1011

\bibitem[{{Perrin} {et~al.}(1998){Perrin}, {Coud{\' e} du Foresto}, {Ridgway},
  {Mariotti}, {Traub}, {Carleton}, \& {Lacasse}}]{PER98}
{Perrin}, G., {Coud{\' e} du Foresto}, V., {Ridgway}, S.~T., {Mariotti}, J.-M.,
  {Traub}, W.~A., {Carleton}, N.~P., \& {Lacasse}, M.~G. 1998, \aap, 331, 619

\bibitem[{{Perrin} {et~al.}(1999){Perrin}, {Coud{\' e} du Foresto}, {Ridgway},
  {Mennesson}, {Ruilier}, {Mariotti}, {Traub}, \& {Lacasse}}]{PER}
{Perrin}, G., {Coud{\' e} du Foresto}, V., {Ridgway}, S.~T., {Mennesson}, B.,
  {Ruilier}, C., {Mariotti}, J.-M., {Traub}, W.~A., \& {Lacasse}, M.~G. 1999,
  \aap, 345, 221

\bibitem[{{Perrin} {et~al.}(2004){Perrin}, {Ridgway}, {Mennesson}, {Cotton},
  {Woillez}, {Verhoelst}, {Schuller}, {Coud{\'e} du Foresto}, {Traub},
  {Millan-Gabet}, \& {Lacasse}}]{PER04}
{Perrin}, G., {Ridgway}, S.~T., {Mennesson}, B., {Cotton}, W.~D., {Woillez},
  J., {Verhoelst}, T., {Schuller}, P., {Coud{\'e} du Foresto}, V., {Traub},
  W.~A., {Millan-Gabet}, R., \& {Lacasse}, M.~G. 2004, \aap, 426, 279

\bibitem[{{Pourbaix} {et~al.}(2002){Pourbaix}, {Platais}, {Detournay},
  {Jorissen}, {Knapp}, \& {Makarov}}]{POU02}
{Pourbaix}, D., {Platais}, I., {Detournay}, S., {Jorissen}, A., {Knapp}, G., \&
  {Makarov}, V.~V. 2002, VizieR Online Data Catalog, 339, 91167

\bibitem[{{Ragland} {et~al.}(2006){Ragland}, {Traub}, {Berger}, {Danchi},
  {Monnier}, {Willson}, {Carleton}, {Lacasse}, {Millan-Gabet}, {Pedretti},
  {Schloerb}, {Cotton}, {Townes}, {Brewer}, {Haguenauer}, {Kern}, {Labeye},
  {Malbet}, {Malin}, {Pearlman}, {Perraut}, {Souccar}, \& {Wallace}}]{RAG06}
{Ragland}, S., {Traub}, W.~A., {Berger}, J.-P., {Danchi}, W.~C., {Monnier},
  J.~D., {Willson}, L.~A., {Carleton}, N.~P., {Lacasse}, M.~G., {Millan-Gabet},
  R., {Pedretti}, E., {Schloerb}, F.~P., {Cotton}, W.~D., {Townes}, C.~H.,
  {Brewer}, M., {Haguenauer}, P., {Kern}, P., {Labeye}, P., {Malbet}, F.,
  {Malin}, D., {Pearlman}, M., {Perraut}, K., {Souccar}, K., \& {Wallace}, G.
  2006, \apj, 652, 650

\bibitem[{{Richichi} \& {Percheron}(2002)}]{CHARM}
{Richichi}, A. \& {Percheron}, I. 2002, \aap, 386, 492

\bibitem[{{Richichi} {et~al.}(2005){Richichi}, {Percheron}, \&
  {Khristoforova}}]{CHARM2}
{Richichi}, A., {Percheron}, I., \& {Khristoforova}, M. 2005, \aap, 431, 773

\bibitem[{{Ridgway} {et~al.}(1992){Ridgway}, {Benson}, {Dyck}, {Townsley}, \&
  {Hermann}}]{RIDG}
{Ridgway}, S.~T., {Benson}, J.~A., {Dyck}, H.~M., {Townsley}, L.~K., \&
  {Hermann}, R.~A. 1992, \aj, 104, 2224

\bibitem[{{Ridgway} {et~al.}(1979){Ridgway}, {Wells}, {Joyce}, \&
  {Allen}}]{RIDG79}
{Ridgway}, S.~T., {Wells}, D.~C., {Joyce}, R.~R., \& {Allen}, R.~G. 1979, \aj,
  84, 247

\bibitem[{{Scholz}(2003)}]{SCH03}
{Scholz}, M. 2003, in Interferometry for Optical Astronomy II, ed.~Wesley A.
  Traub, SPIE Proc., 4838, 163

\bibitem[{{Scholz} \& {Wood}(2000)}]{SW}
{Scholz}, M. \& {Wood}, P.~R. 2000, \aap, 362, 1065

\bibitem[{{Sloan} \& {Price}(1998)}]{SLOANPRICE}
{Sloan}, G.~C. \& {Price}, S.~D. 1998, \apjs, 119, 141

\bibitem[{{Smith} {et~al.}(2002){Smith}, {Leisawitz}, {Castelaz}, \&
  {Luttermoser}}]{SMITH02}
{Smith}, B.~J., {Leisawitz}, D., {Castelaz}, M.~W., \& {Luttermoser}, D. 2002,
  \aj, 123, 948

\bibitem[{{Tej} {et~al.}(1999){Tej}, {Chandrasekhar}, {Ashok}, {Ragland},
  {Richichi}, \& {Stecklum}}]{TEJ99}
{Tej}, A., {Chandrasekhar}, T., {Ashok}, N.~M., {Ragland}, S., {Richichi}, A.,
  \& {Stecklum}, B. 1999, \aj, 117, 1857

\bibitem[{{Tej} {et~al.}(2003){Tej}, {Lan{\c c}on}, {Scholz}, \& {Wood}}]{TLSW}
{Tej}, A., {Lan{\c c}on}, A., {Scholz}, M., \& {Wood}, P.~R. 2003, \aap, 412,
  481

\bibitem[{{Thompson} {et~al.}(2002){Thompson}, {Creech-Eakman}, \& {van
  Belle}}]{THOM}
{Thompson}, R.~R., {Creech-Eakman}, M.~J., \& {van Belle}, G.~T. 2002, \apj,
  577, 447

\bibitem[{{Tuthill} {et~al.}(2000{\natexlab{a}}){Tuthill}, {Danchi}, {Hale},
  {Monnier}, \& {Townes}}]{TUT00b}
{Tuthill}, P.~G., {Danchi}, W.~C., {Hale}, D.~S., {Monnier}, J.~D., \&
  {Townes}, C.~H. 2000{\natexlab{a}}, \apj, 534, 907

\bibitem[{{Tuthill} {et~al.}(1995){Tuthill}, {Haniff}, \& {Baldwin}}]{TUT95}
{Tuthill}, P.~G., {Haniff}, C.~A., \& {Baldwin}, J.~E. 1995, \mnras, 277, 1541

\bibitem[{{Tuthill} {et~al.}(1999){Tuthill}, {Haniff}, \& {Baldwin}}]{TUT99}
---. 1999, \mnras, 306, 353

\bibitem[{{Tuthill} {et~al.}(2002){Tuthill}, {Monnier}, {Danchi}, {Hale}, \&
  {Townes}}]{TUT02}
{Tuthill}, P.~G., {Monnier}, J.~D., {Danchi}, W.~C., {Hale}, D.~D.~S., \&
  {Townes}, C.~H. 2002, \apj, 577, 826

\bibitem[{{Tuthill} {et~al.}(2000{\natexlab{b}}){Tuthill}, {Monnier}, {Danchi},
  {Wishnow}, \& {Haniff}}]{TUT00}
{Tuthill}, P.~G., {Monnier}, J.~D., {Danchi}, W.~C., {Wishnow}, E.~H., \&
  {Haniff}, C.~A. 2000{\natexlab{b}}, \pasp, 112, 555

\bibitem[{{van Belle} {et~al.}(1996){van Belle}, {Dyck}, {Benson}, \&
  {Lacasse}}]{VANB}
{van Belle}, G.~T., {Dyck}, H.~M., {Benson}, J.~A., \& {Lacasse}, M.~G. 1996,
  \aj, 112, 2147

\bibitem[{{van Belle} {et~al.}(2002){van Belle}, {Thompson}, \&
  {Creech-Eakman}}]{VANB02}
{van Belle}, G.~T., {Thompson}, R.~R., \& {Creech-Eakman}, M.~J. 2002, \aj,
  124, 1706

\bibitem[{{Whitelock} {et~al.}(2000){Whitelock}, {Marang}, \& {Feast}}]{WHI}
{Whitelock}, P., {Marang}, F., \& {Feast}, M. 2000, \mnras, 319, 728

\bibitem[{{Wood} \& {Karovska}(2006)}]{WOO06}
{Wood}, B.~E. \& {Karovska}, M. 2006, \apj, 649, 410

\bibitem[{{Wood} {et~al.}(1999){Wood}, {Alcock}, {Allsman}, {Alves}, {Axelrod},
  {Becker}, {Bennett}, {Cook}, {Drake}, {Freeman}, {Griest}, {King}, {Lehner},
  {Marshall}, {Minniti}, {Peterson}, {Pratt}, {Quinn}, {Stubbs}, {Sutherland},
  {Tomaney}, {Vandehei}, \& {Welch}}]{woodM}
{Wood}, P.~R., {Alcock}, C., {Allsman}, R.~A., {Alves}, D., {Axelrod}, T.~S.,
  {Becker}, A.~C., {Bennett}, D.~P., {Cook}, K.~H., {Drake}, A.~J., {Freeman},
  K.~C., {Griest}, K., {King}, L.~J., {Lehner}, M.~J., {Marshall}, S.~L.,
  {Minniti}, D., {Peterson}, B.~A., {Pratt}, M.~R., {Quinn}, P.~J., {Stubbs},
  C.~W., {Sutherland}, W., {Tomaney}, A., {Vandehei}, T., \& {Welch}, D.~L.
  1999, in IAU Symp. 191: Asymptotic Giant Branch Stars, eds.~T.~ Le Bertre,
  A.~Lebre and C.~Waelkens, 151

\bibitem[{{Woodruff} {et~al.}(2004){Woodruff}, {Eberhardt}, {Driebe},
  {Hofmann}, {Ohnaka}, {Richichi}, {Schertl}, {Sch{\" o}ller}, {Scholz},
  {Weigelt}, {Wittkowski}, \& {Wood}}]{WOO}
{Woodruff}, H.~C., {Eberhardt}, M., {Driebe}, T., {Hofmann}, K., {Ohnaka}, K.,
  {Richichi}, A., {Schertl}, D., {Sch{\" o}ller}, M., {Scholz}, M., {Weigelt},
  G., {Wittkowski}, M., \& {Wood}, P. 2004, \aap, 421, 703

\bibitem[{{Young} {et~al.}(2000){Young}, {Baldwin}, {Boysen}, {Haniff},
  {Pearson}, {Rogers}, {St-Jacques}, {Warner}, \& {Wilson}}]{YOU00}
{Young}, J.~S., {Baldwin}, J.~E., {Boysen}, R.~C., {Haniff}, C.~A., {Pearson},
  D., {Rogers}, J., {St-Jacques}, D., {Warner}, P.~J., \& {Wilson}, D.~M.~A.
  2000, \mnras, 318, 381

\end{thebibliography}

\clearpage

\end{document}